\documentclass[a4paper,11pt]{article}
\pdfoutput=1
\usepackage{jheppub,MnSymbol}

\def\ms{\overline{\rm MS}}

\def\disg{{\rm DIS}_\gamma}
\def\f2y{F_2^\gamma}

\def\p{I\!\!P}
\def\Pomeron{I\!\!P}
\newcommand{\pom}{{I\!\!P}}
\newcommand{\reg}{{I\!\!R}}
\newcommand{\xpom}{x_{_{I\!\!P}}}
\newcommand{\alphapom}{\alpha_{_{\rm I\!P}}}

\def\lr{\left( }
\def\rr{\right) }

\newcommand{\beq}{\begin{equation}}
\newcommand{\eeq}{\end{equation}}
\newcommand{\bea}{\begin{eqnarray}}
\newcommand{\eea}{\end{eqnarray}}

\preprint{MS-TP-20-18}

\title{Diffractive dijet photoproduction at the EIC}

\author[a]{Vadim Guzey}
\author[b]{and Michael Klasen}

\affiliation[a]{National Research Center ``Kurchatov Institute'', Petersburg
 Nuclear Physics Institute (PNPI), Gatchina, 188300, Russia}
\affiliation[b]{Institut f\"ur Theoretische Physik, Westf\"alische Wilhelms-Universit\"at M\"unster, Wilhelm-Klemm-Stra\ss{}e 9, 48149 M\"unster, Germany}

\emailAdd{guzey\_va@nrcki.pnpi.ru}
\emailAdd{michael.klasen@uni-muenster.de}

\abstract{We present a first, detailed study of diffractive dijet photoproduction at
  the recently approved electron-ion collider (EIC) at BNL. Apart from establishing
  the kinematic reaches for various beam types, energies and kinematic cuts, we make
  precise predictions at next-to-leading order (NLO) of QCD in the most important
  kinematic variables. We show that the EIC will provide new and more precise information
  on the diffractive parton density functions (PDFs) in the pomeron than previously
  obtained at HERA, illuminate the still disputed mechanism of global vs.\ only resolved-photon
  factorization breaking, and provide access to a completely new quantity, i.e.\
  nuclear diffractive PDFs.
}


\begin{document}
\maketitle
\flushbottom

\section{Introduction}
\label{sec:1}

From 1992 to 2007, hadron-electron collisions at DESY's circular installation
HERA provided a wealth of data and information on the strong interaction and
the partonic structure of the proton, not only in deep-inelastic scattering
(DIS), where the virtuality of the exchanged photon $Q^2\gg 1$ GeV$^2$ is
large, but also in photoproduction, where $Q^2\leq 1$ GeV$^2$. While electrons
(or positrons) of energy 27.5 GeV mostly collided with protons of first 820,
then 920 GeV, the last months of operation were dedicated to lower proton
beam energies of 575 and 460 GeV in view of better access to the longitudinal
structure function $F_L$ and therefore the gluon dynamics at small momentum
fractions $x$. While inclusive DIS precisely pinned down the quark parton
distribution functions (PDFs), jets (which for dijet invariant masses larger
than 16 GeV make up $10-20$\% of the inclusive DIS cross section) and
photoproduction provided additional constraints on the running of the QCD
coupling constant $\alpha_s$ and the gluon PDF in the proton
\cite{Newman:2013ada,Klasen:2002xb}.

The two general purpose detectors H1 and ZEUS were supplemented with forward
taggers to identify diffractive processes $ep\to eXY$, which -- rather
surprisingly -- accounted for a substantial fraction ($10-15$\%) of all
events. In DIS, where QCD factorization was proven to hold
\cite{Collins:1997sr}, they could be interpreted in terms of diffractive
structure functions. Under the additional assumption of Regge factorization
\cite{Ingelman:1984ns}, the flux of pomeron ($\p$) and higher Regge
trajectories like the reggeon ($\reg$) can be parametrized, pomeron PDFs could
be extracted \cite{Aktas:2006hy,Aktas:2007bv,Chekanov:2009aa}
and their universality tested in other types of collisions. It turned
out that in hadron-hadron collisions at the Tevatron and the LHC,
factorization was broken by a factor of 0.1 \cite{Affolder:2000vb} to 0.025
\cite{Sirunyan:2020ifc}, depending on the collision energy, the order of
perturbative QCD calculations and simplifying assumptions about the
non-diffractive structure function \cite{Klasen:2009bi} and in qualitative
agreement with calculations based on multi-pomeron exchanges in a two-channel
eikonal model \cite{Khoze:2000wk}.

For dijet photoprodution, these calculations
suggest that factorization should hold for the DIS-like direct-photon
processes, but be broken by a factor of 0.34 for the resolved-photon
processes, where the photon fluctuates before the hard interaction into
$q\bar{q}$ pairs and their vector-meson dominated (VMD) bound states
\cite{Kaidalov:2003xf}. It is, however, well-known that at next-to-leading
order (NLO) of QCD \cite{Klasen:1994bj} and beyond \cite{Klasen:2013cba}
direct and resolved processes are connected through the factorization of
collinear initial-state singularities, which to preserve factorization-scale
independence should also be suppressed \cite{Klasen:2005dq}. Also, the H1
\cite{Aktas:2007hn} and ZEUS
\cite{Chekanov:2007rh} data can be described by not only suppressing the
resolved (and direct initial-state) contribution, but also by a global
suppression factor of 0.42 to 0.71, even though this factor then depends on
the transverse jet momentum and is subject to large theoretical uncertainties
from scale variations and hadronization corrections \cite{Klasen:2008ah}.
It is also possible that the resolved-photon suppression factor depends on
the parton flavor \cite{Guzey:2016awf}. Elucidating the mechanism of
factorization breaking in dijet photoproduction is therefore one of the
important desiderata of the HERA program.

Since there is currently no electron-hadron collider in operation, new
experimental information can in the short term only be obtained from
ultraperipheral collisions (UPCs) at the LHC \cite{Baltz:2007kq}, which have
contributions from both photoproduction \cite{Guzey:2018dlm} and diffraction
\cite{Guzey:2016tek}. Dijet photoproduction at the LHC might even provide
novel constraints on nuclear PDFs \cite{Guzey:2019kik} or first information
on the yet unknown diffractive nuclear PDFs \cite{Frankfurt:2011cs,%
Guzey:2016tek}. In the medium term, the recently approved electron-ion
collider (EIC) at BNL \cite{Accardi:2012qut} has the potential for detailed
studies of jets in DIS
\cite{Boer:2016fqd,Klasen:2017kwb, Boughezal:2018azh,Liu:2018trl,%
Hatta:2019ixj,Mantysaari:2019hkq} and photoproduction 
\cite{Klasen:2018gtb,Chu:2017mnm,
Aschenauer:2019uex} in the clean environment of an electron-nucleus
collider, which was planned for Run 3 at HERA, but never implemented.

In this paper, we explore in detail the EIC potential for diffractive dijet
photoproduction. We begin by reviewing in Sec.\ \ref{sec:2} our analytical
approach
based on the factorization of hadronic and partonic cross sections,
the extraction of diffractive PDFs from HERA, our NLO QCD calculation of the
partonic dijet cross section, and theoretical models for factorization
breaking. Section \ref{sec:3} contains a large variety of results for
diffraction on protons, starting with NLO QCD predictions for the EIC with
colliding beams of 21 GeV electrons and 100 GeV protons and the
corresponding $K$ factors as well as studies of the scale evolution of the
pomeron PDFs, the range of cross sections predicted from different HERA fits
of the diffractive PDFs, of the cross section with a larger range in the
longitudinal pomeron momentum fraction $x_{\p}$ and of the corresponding
increase of the sub-leading reggeon contribution. We then demonstrate the
advantage of a higher proton beam energy of 275 GeV and conclude this section
with numerical predictions based on the different approaches to factorization
breaking. In Sec.\ \ref{sec:4} we address the diffraction on nuclei. We start
by reviewing the theoretical definition of nuclear diffractive PDFs and the
leading-twist model of nuclear shadowing, before we make numerical predictions
at NLO for diffractive dijet photoproduction on various nuclei and discuss
again the different approaches to factorization breaking. Our conclusions
are given in Sec.\ \ref{sec:5}.

\section{Analytical approach}
\label{sec:2}

At the EIC, like at HERA, electrons $e$ of four-momentum $k$ will collide with
protons $p$ of four-momentum $P$ at a squared center-of-mass system (CMS)
energy $S=(k+P)^2$. For nuclei, the relevant quantity is the squared CMS energy
per nucleon and is typically (i.e.\ for heavy nuclei) smaller by about a factor
of 
$Z/A \approx 0.4$, where $Z$ is the nucleus charge and $A$ is the number of nucleons.
In photoproduction, the virtuality $Q^2 = -q^2=-(k-k')^2$ of the
radiated photon $\gamma$ is small (typically less than $Q_{\max}^2=0.01-1$
GeV$^2$), and its spectrum can be described in the improved
Weizs\"acker-Williams approximation \cite{Frixione:1993yw}
\beq
 f_{\gamma /e}(y) = \frac{\alpha }{2\pi} \left[\frac{1+(1-y)^2}{y} \ln 
\frac{Q^2_{\max}(1-y)}{m_e^2y^2}
 +  2m_e^2y\lr\frac{1-y}{m_e^2y^2} -\frac{1}{Q^2_{\max}}\rr\right].
\eeq
Here, $\alpha$ is the electromagnetic fine structure constant,
$k'$ is the four-momentum of the scattered electron, $y = (qP)/(kP)$
is its longitudinal momentum transfer and $m_e$ its mass.

Diffractive processes are characterized by the presence of a large rapidity
gap between the central hadronic system $X$ and the forward-going hadronic
system $Y$ with four-momentum $p_Y$, low mass $M_Y$ (typically a proton that
remained intact or a proton plus low-lying nucleon resonances), small
four-momentum transfer $t=(P-p_Y)^2$, and small longitudinal momentum transfer
$x_{\p} = q(P-p_Y)/(qP)$  (see Fig.\ \ref{fig:1}).
%
{\begin{figure}\centering
 \epsfig{file=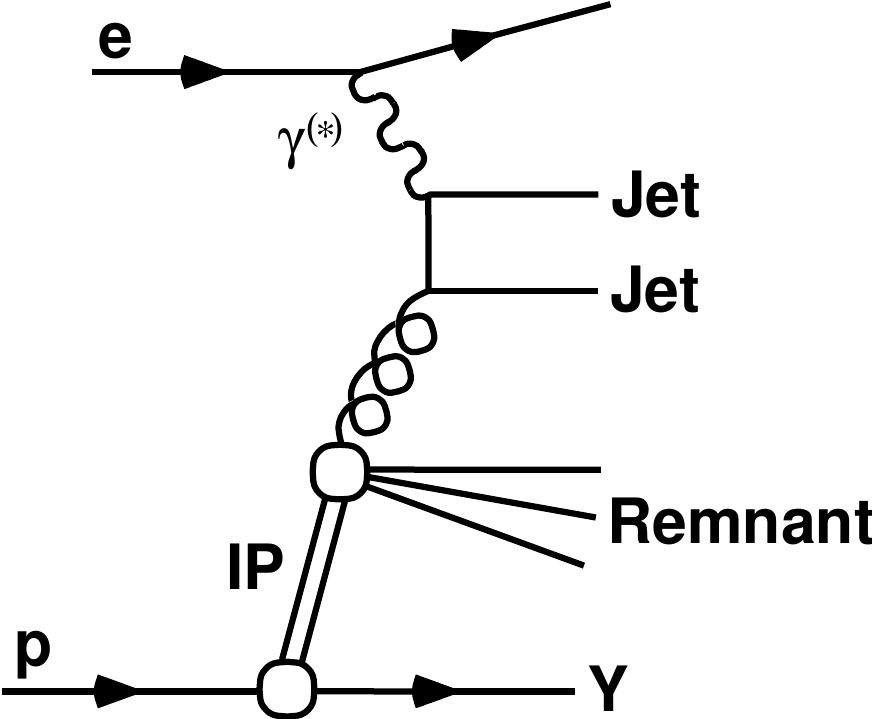,width=0.46\textwidth}
 \epsfig{file=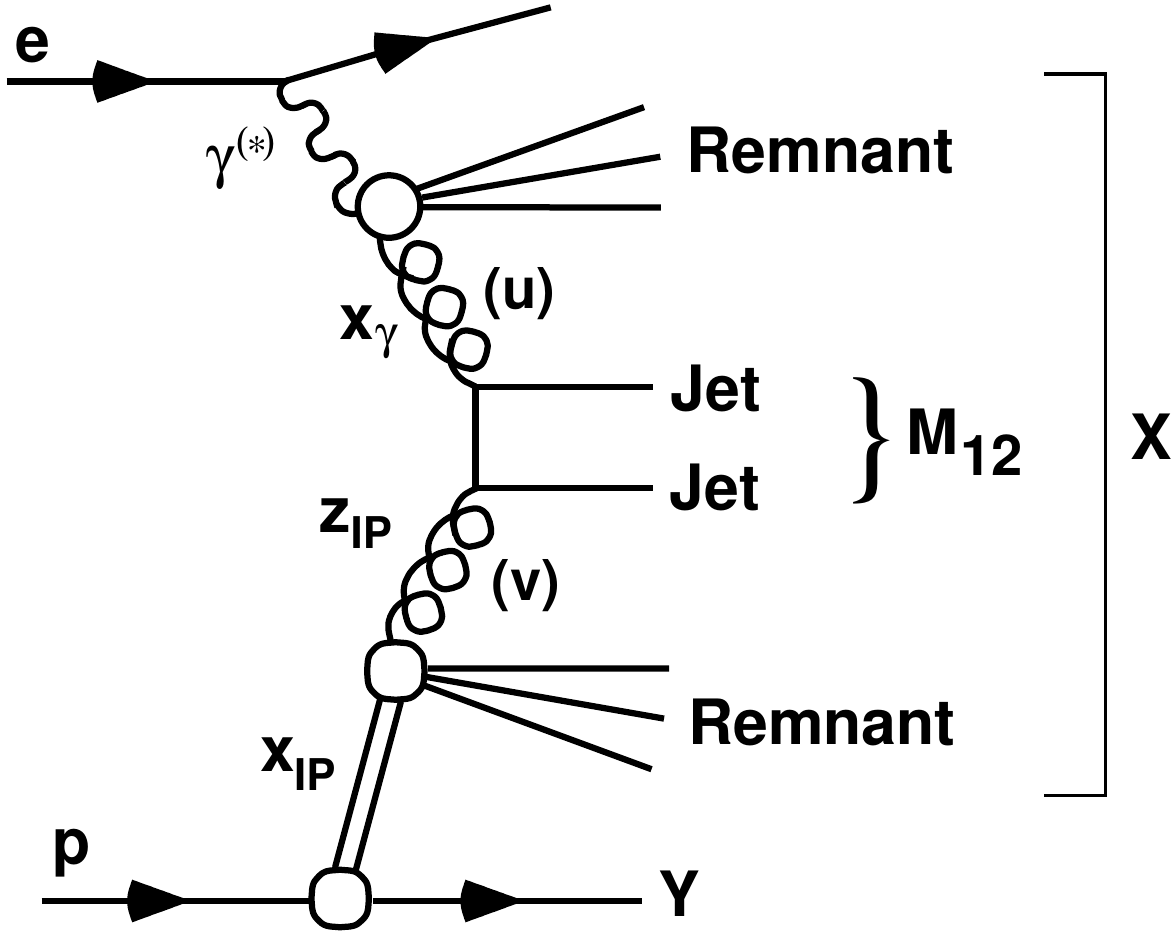,width=0.49\textwidth}
\caption{Diffractive production of dijets with invariant mass
  $M_{12}$ in direct (left) and resolved (right) photon-pomeron collisions,
  leading to the production of one or two additional remnant jets.
  The hadronic systems $X$ and $Y$ are separated by the largest rapidity
  gap in the final state.}
\label{fig:1}
\end{figure}}
%

In dijet photoproduction, the system $X$ contains (at least) two hard jets with
transverse momenta $p_{T1,2}$, rapidities $\eta_{1,2}$ and invariant mass
$M_{12}$, as well as remnant jets from the diffractive exchange, dominated by
the pomeron $\p$ as the lowest-lying Regge trajectory, and 
from
the photon, when
the latter does not interact directly with the proton or nucleus, but first
resolves into its partonic constituents. Assuming both QCD and Regge
factorization, the cross section for the reaction $e+p \rightarrow e+
2~{\rm jets}+X'+Y$ can then be calculated through
\beq
  d\sigma =
  \sum_{a,b} \int\!dy \int\!dx_{\gamma}\!\int\!dt\!\int\!dx_{\p}\!\int\!dz_{\p}
  f_{\gamma/e}(y) f_{a/\gamma }(x_\gamma,M^2_{\gamma }) f_{\p/p}(x_{\p},t)
  f_{b/\p}(z_{\p},M_{\p}^2) d\hat{\sigma}_{ab}^{(n)}.
  \label{eq:2.2}
\eeq
The $\xpom$ dependence is parameterized using a flux factor motivated by Regge
theory,
\begin{eqnarray}
f_{\pom/p}(\xpom, t) = A_\pom \cdot 
\frac{e^{B_\pom t}}{\xpom^{2\alpha_\pom (t)-1}} \ ,
\label{eq:fluxfac}
\end{eqnarray}
where the pomeron trajectory is assumed to be linear, $\alpha_\pom (t)=
\alpha_\pom (0) + \alpha_\pom^\prime t$, and the parameters $B_\pom$ and
$\alphapom^\prime$ and their uncertainties are obtained from fits to H1
diffractive DIS data \cite{Aktas:2006hy}. The longitudinal momentum fractions
of the parton $a$ in the photon $x_\gamma$ and of the parton $b$ in the
pomeron $z_{\p}$ can be experimentally determined from the two observed
leading jets through
\bea
x_{\gamma}^{\rm obs} = {p_{T1}\,e^{-\eta_1}+p_{T2}\,e^{-\eta_2}\over 2yE_e}
&\ {\rm and}\ &
z_{\p}^{\rm obs} = {p_{T1}\,e^{\eta_1}+p_{T2}\,e^{\eta_2}\over 2x_{\p}E_p}.
\eea
$M_{\gamma}$ and $M_{\p}$ are the factorization scales at the respective
vertices, and $d\hat{\sigma}_{ab}^{(n)}$ is the cross section for the
production of an $n$-parton final state from two initial partons $a$ and $b$.
It is calculated in NLO in $\alpha_s(\mu)$ \cite{Klasen:1994bj},
as are the PDFs of the photon and
the pomeron. For the former, we use the GRV NLO parametrization, which we
transform from the $\disg$ to the $\ms$ scheme \cite{Gluck:1991jc}. Our
default choice for the diffractive PDFs is H1 2006 Fit B \cite{Aktas:2006hy},
which includes proton dissociation up to masses of $M_Y<1.6$ GeV and is
integrated up to $|t|<1$ GeV$^2$ and $x_{\p}<0.03$. We identify the
factorization scales $M_{\gamma}$, $M_{\p}$ and the renormalization scale
$\mu$ with the average transverse momentum $\bar{p}_T=(p_{T1}+p_{T2})/2$
\cite{Klasen:2008ah}.

\section{Diffraction on protons}
\label{sec:3}

In this first numerical section, we focus on electron-proton collisions
at the EIC with an electron beam energy of $E_e=21$ GeV and a proton
beam energy of $E_p=100$ GeV, which will in the next section also be used
as the beam energy per nucleon for electron-nucleus collisions. We assume
detectors that have the same kinematic acceptance as H1 for diffractive
events, i.e.\ the capability to identify a large rapidity gap and/or a
leading proton in a Roman pot spectrometer. We also allow for proton
dissociation up to masses of $M_Y<$ 1.6 GeV, a four-momentum
transfer of $|t|<1$ GeV$^2$ and a longitudinal momentum transfer of
$x_{\p}<0.03$. Photoproduction events are assumed to be selected with
\mbox{(anti-)}tagged electrons and photon virtualities up to $Q^2<0.1$ GeV$^2$,
assuming full kinematic coverage of the longitudinal momentum transfer
$0<y<1$ from the electron to the photon. 

Jets are defined with the anti-$k_T$ algorithm and a distance parameter $R=1$,
where at NLO jets contain at most two partons \cite{Cacciari:2008gp}. Given
the limited EIC energy and experience from HERA, we assume that the detectors
can identify jets above relatively low tranverse energies of $p_{T1}>5$ GeV
(leading jet) and $p_{T2}>4.5$ GeV (subleading jet). This will, however,
require a good resolution of the hadronic jet energy scale and subtraction of
the underlying event. The latter will also be important to avoid large
hadronization corrections of the partons, which are particularly prominent at
large $x_{\gamma}$ and have so far obscured the interpretation of the observed
factorization breaking. Note also that asymmetric jet $p_T$ cuts allow 
one
to
avoid an enhanced
sensitivity to soft radiation \cite{Klasen:1995xe}. Rapidities are a priori
accepted in the range $\eta_{1,2}\in[-4;4]$. We find, however, that in
diffractive photoproduction most jets are central and have an average rapidity
$\bar{\eta}=(\eta_1+\eta_1)/2\in[-1.5;0]$. This range is enlarged to $[-1.5;1]$
at higher proton beam energy or for a larger range in $x_{\p}$, see below.

\subsection{NLO QCD predictions for the EIC}

%
{\begin{figure}\centering
 \epsfig{file=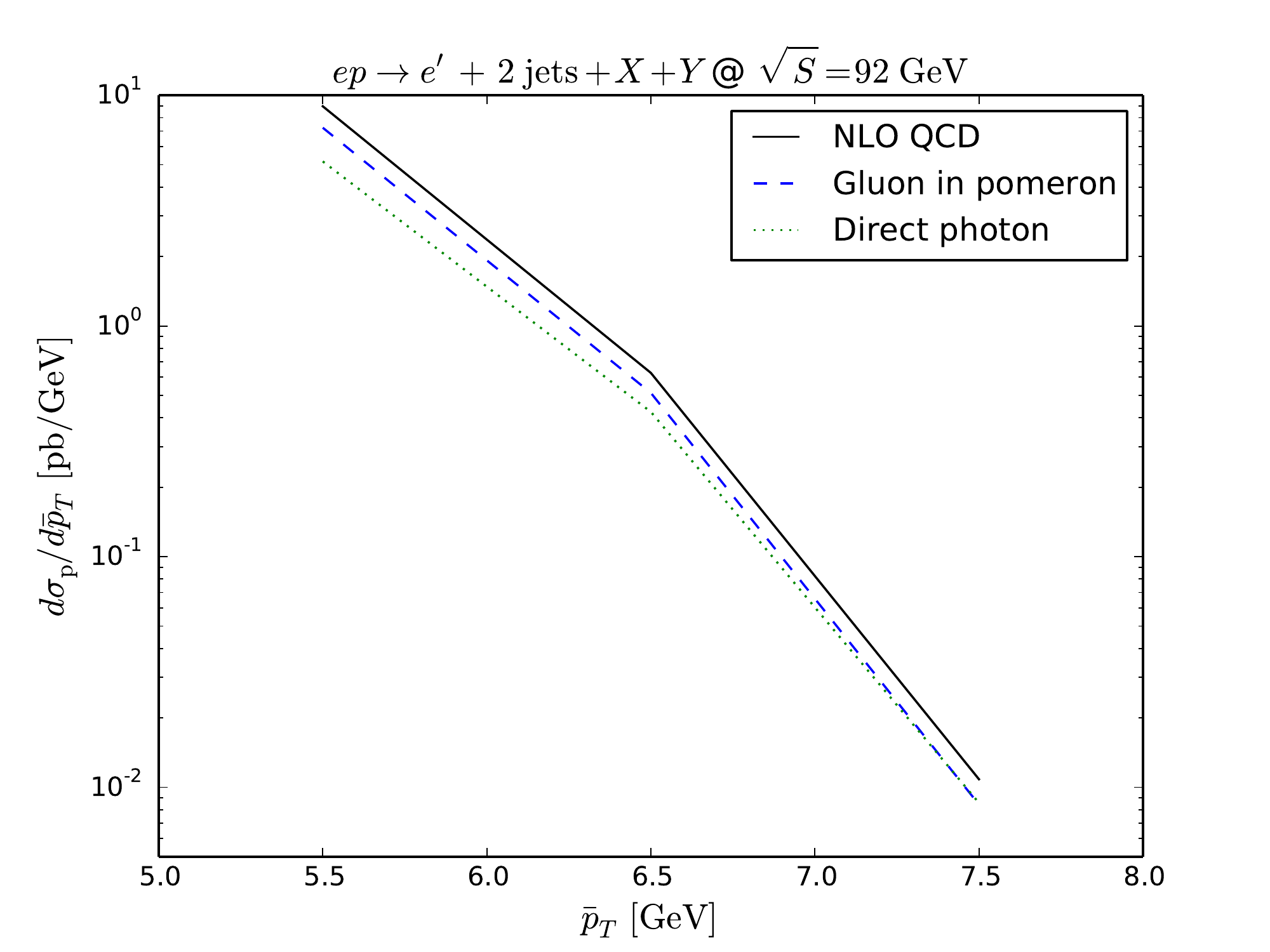,width=0.49\textwidth}
 \epsfig{file=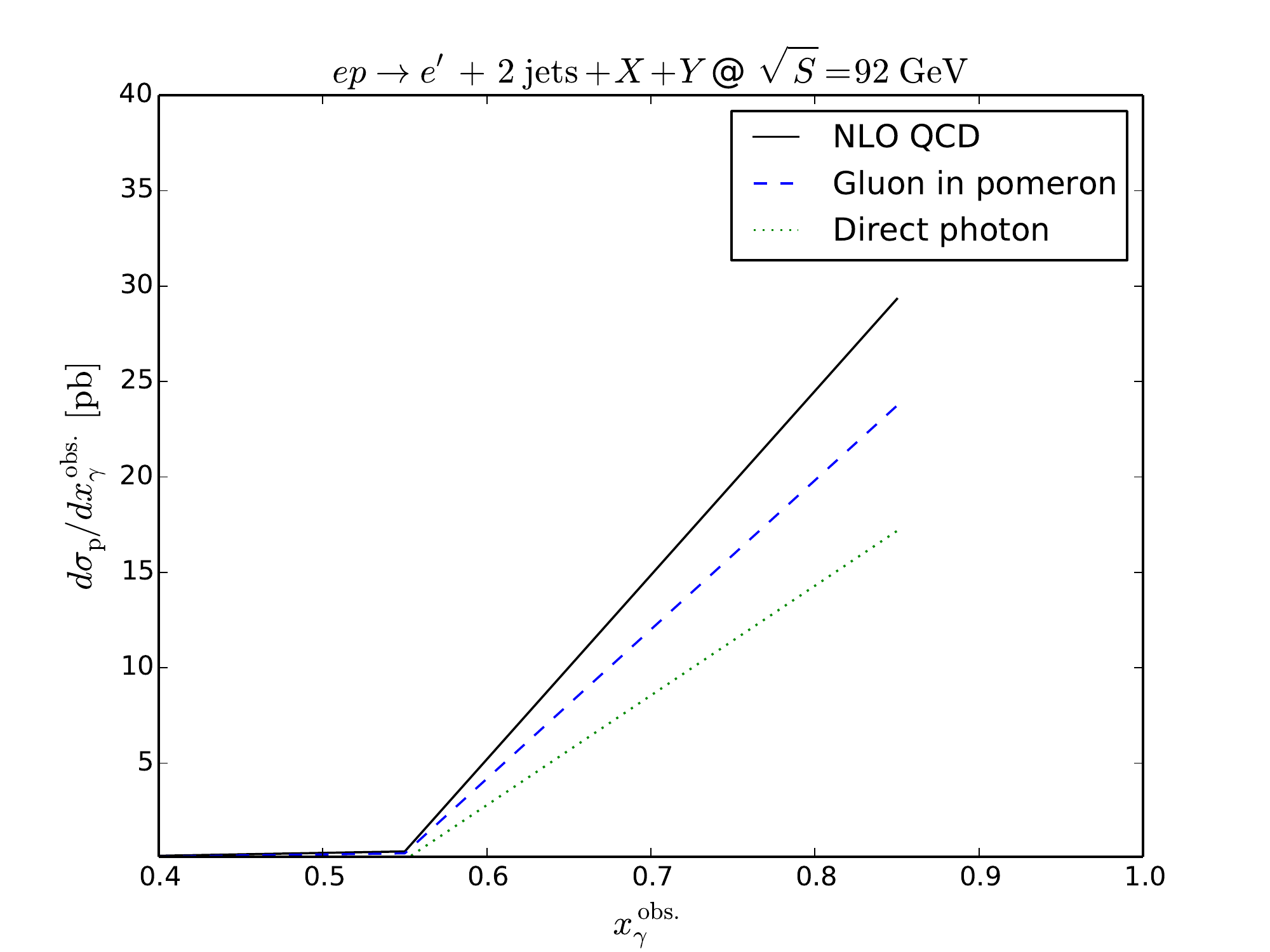,width=0.49\textwidth}
 \epsfig{file=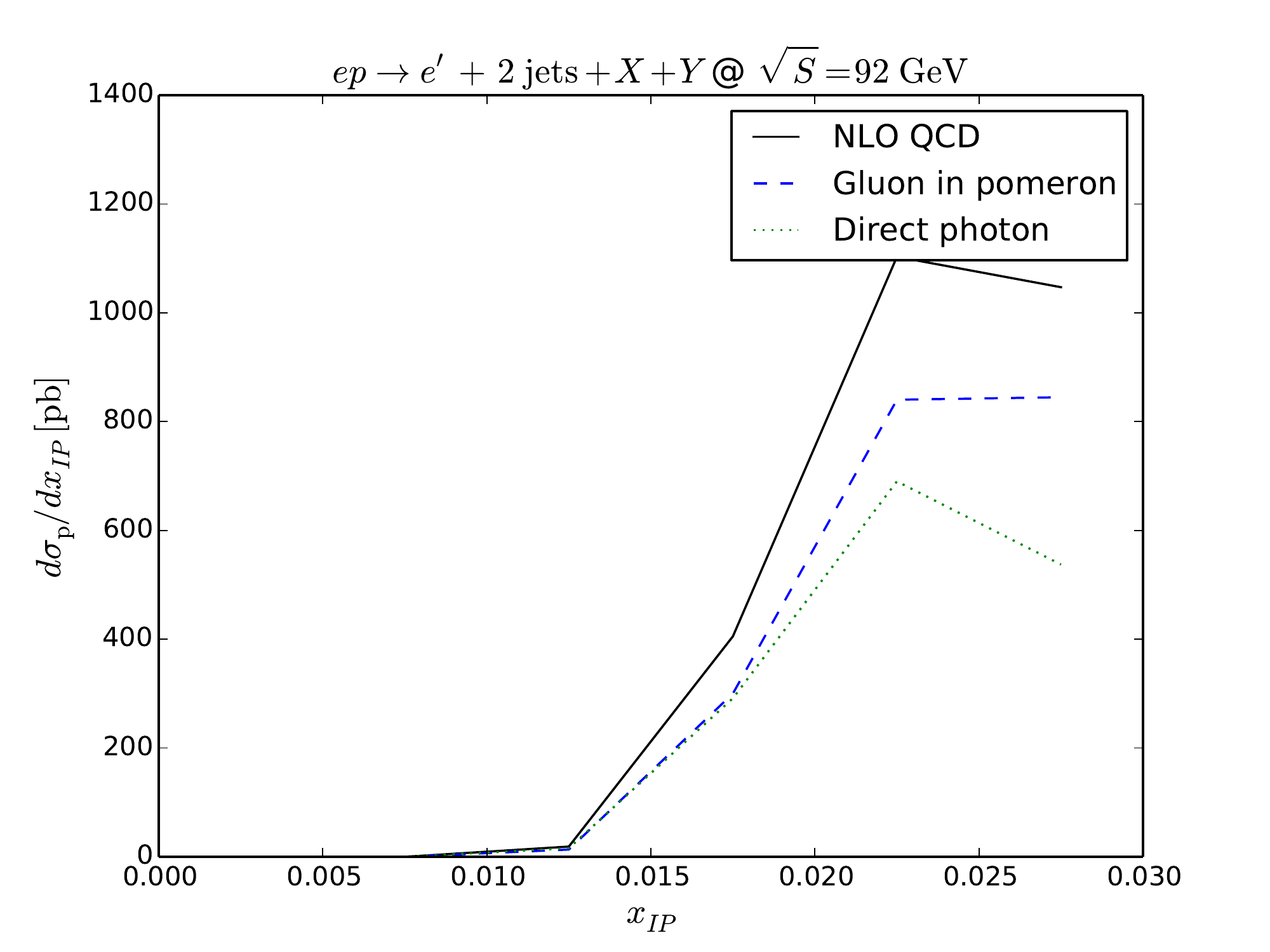,width=0.49\textwidth}
 \epsfig{file=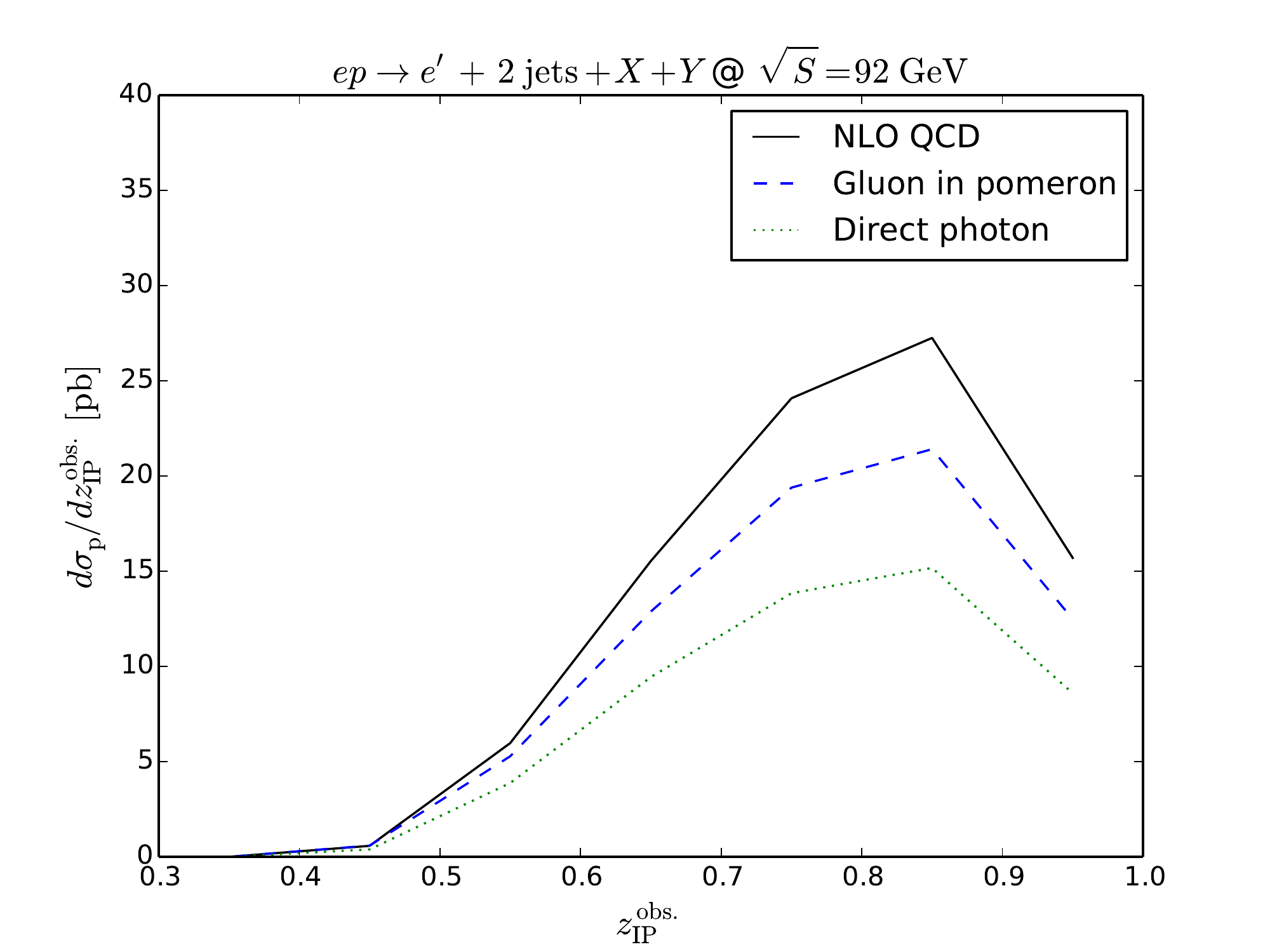,width=0.49\textwidth}
 \caption{NLO QCD cross sections for diffractive dijet photoproduction at the
   EIC in our default set-up. Shown are distributions in the jet average
   transverse momentum (top left) as well as the (observed) longitudinal
   momentum fractions of the photon (top right), the pomeron (bottom left)
   and the partons in the pomeron (bottom right). In addition to the total
   cross section (full black) we also show the contributions from gluons in the
   pomeron (dashed blue) and direct photons (dotted green curves).}
\label{fig:2}
\end{figure}}
%
In Fig.\ \ref{fig:1} we show our NLO QCD cross sections for diffractive dijet
photoproduction at the EIC in this default set-up, i.e.\ at a CMS energy of
92 GeV. The distribution in the jet average transverse momentum (top left)
extends only to 8 GeV, while at HERA with its larger CMS energy of $300-320$
GeV it extended to about 15 GeV. Consequently, the total cross sections (full
black curves) are dominated by contributions from direct photons (dotted green
curves) and point-like quark-antiquark pairs, as
one can also see from the accessible range in $x_{\gamma}^{\rm obs.}> 0.5$
(top right). It will thus not be easy at this CMS energy to distinguish global
factorization breaking from a breaking in only the resolved-photon
contribution. Due to the limited available energy, the cross section also
requires the largest longitudinal momentum fraction allowed by the kinematic
cut of the proton to the pomeron (bottom left) and is dominated by large
momentum fractions of the partons (mostly gluons, dashed blue curves)
in the pomeron (bottom right).

\subsection{$K$-factors}

For detailed event studies with Monte Carlo generators in leading order (LO)
QCD, it is useful to know the impact of the NLO corrections. Therefore we show
in Fig.\ \ref{fig:2} the $K$-factors, i.e.\ the ratio of NLO over LO cross%
%
{\begin{figure}\centering
 \epsfig{file=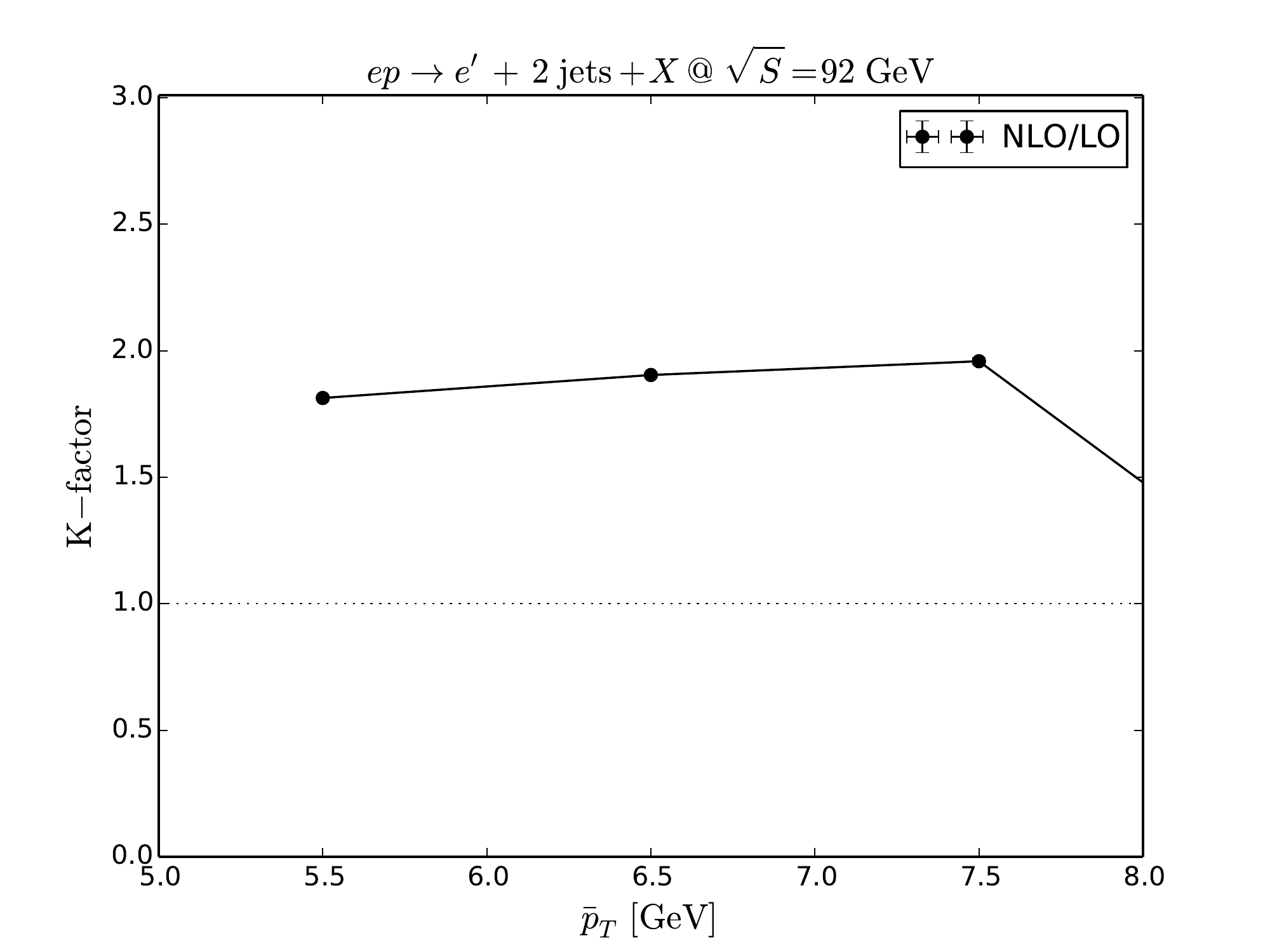,width=0.49\textwidth}
 \epsfig{file=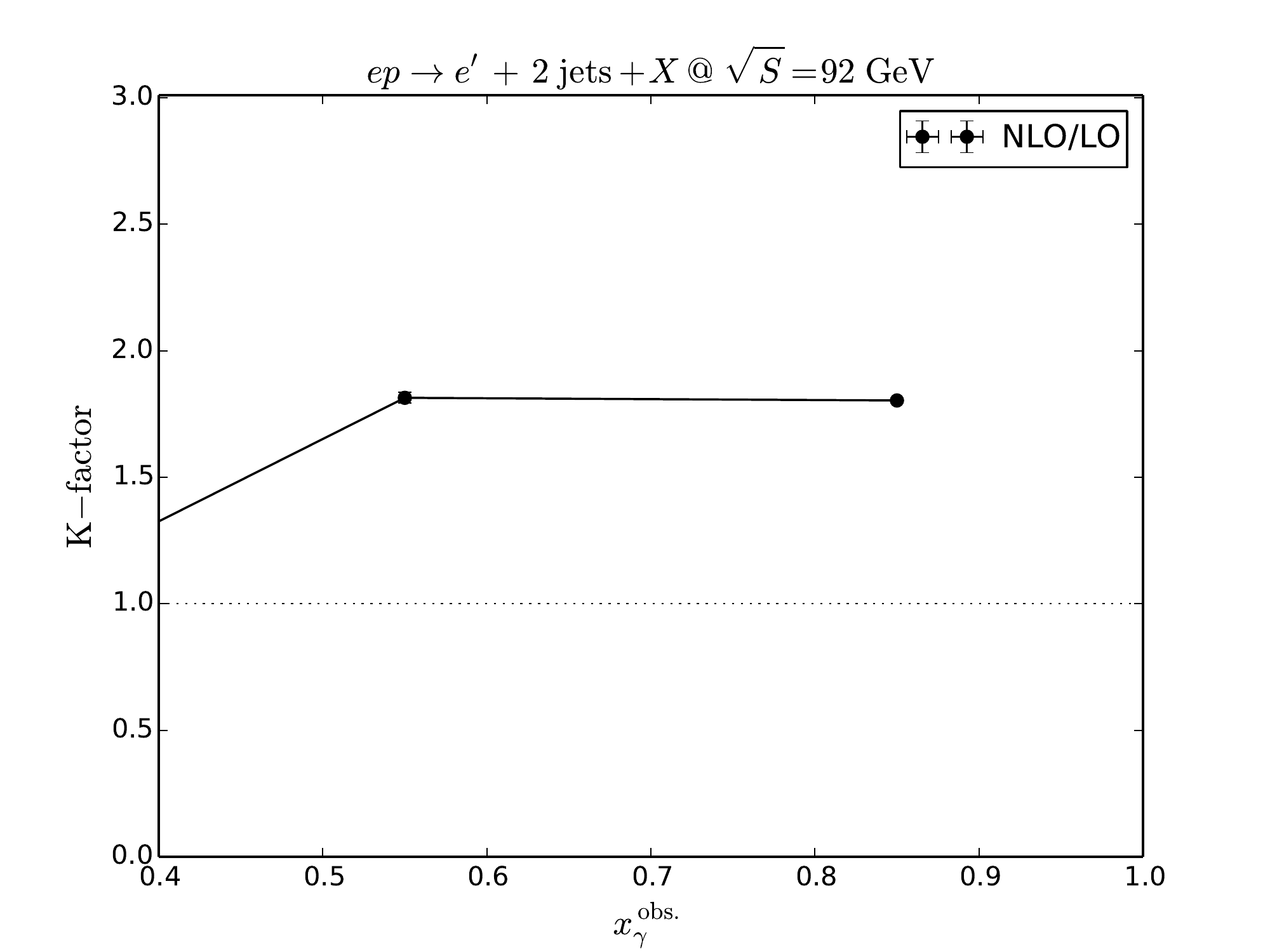,width=0.49\textwidth}
 \epsfig{file=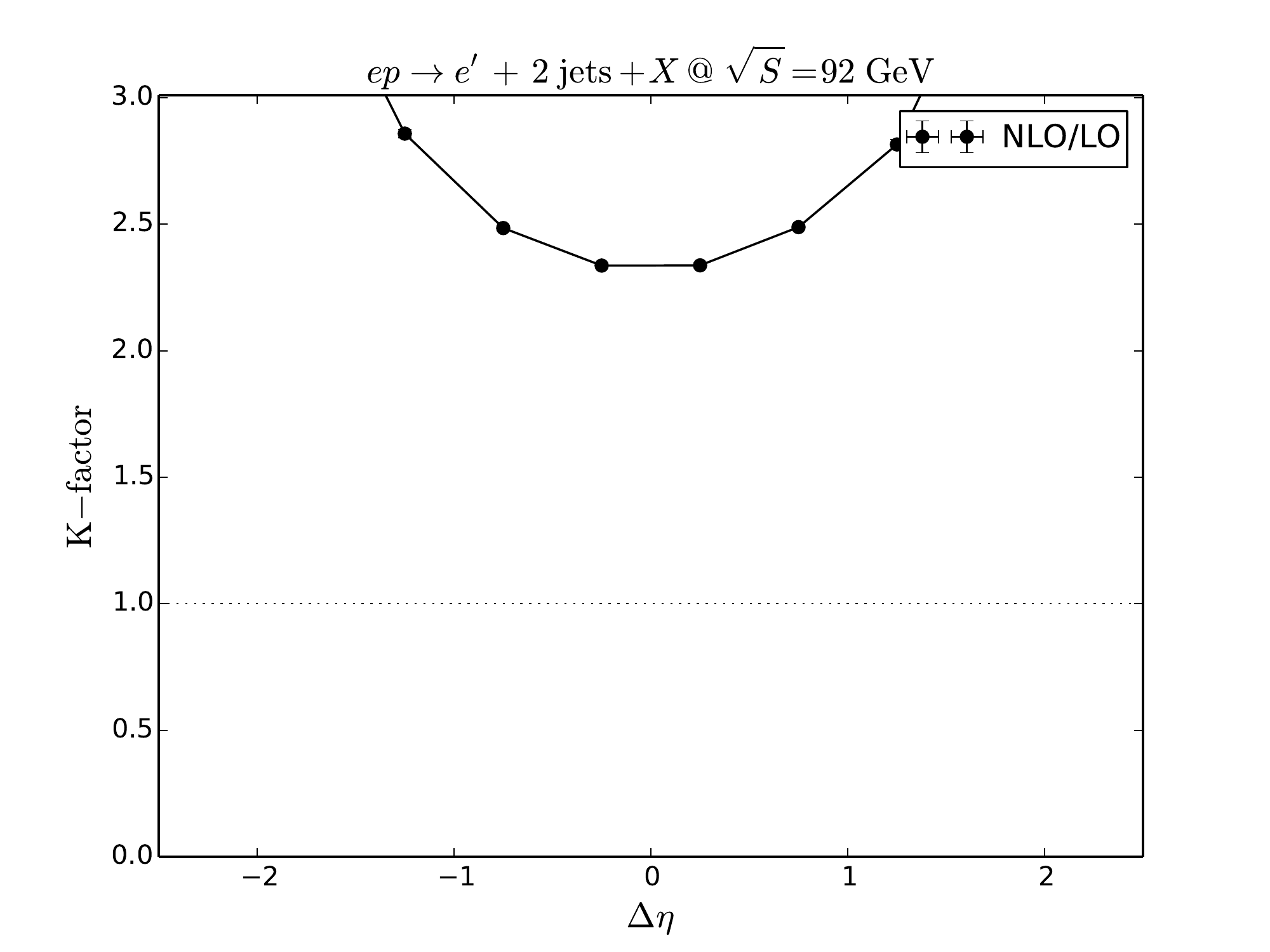,width=0.49\textwidth}
 \epsfig{file=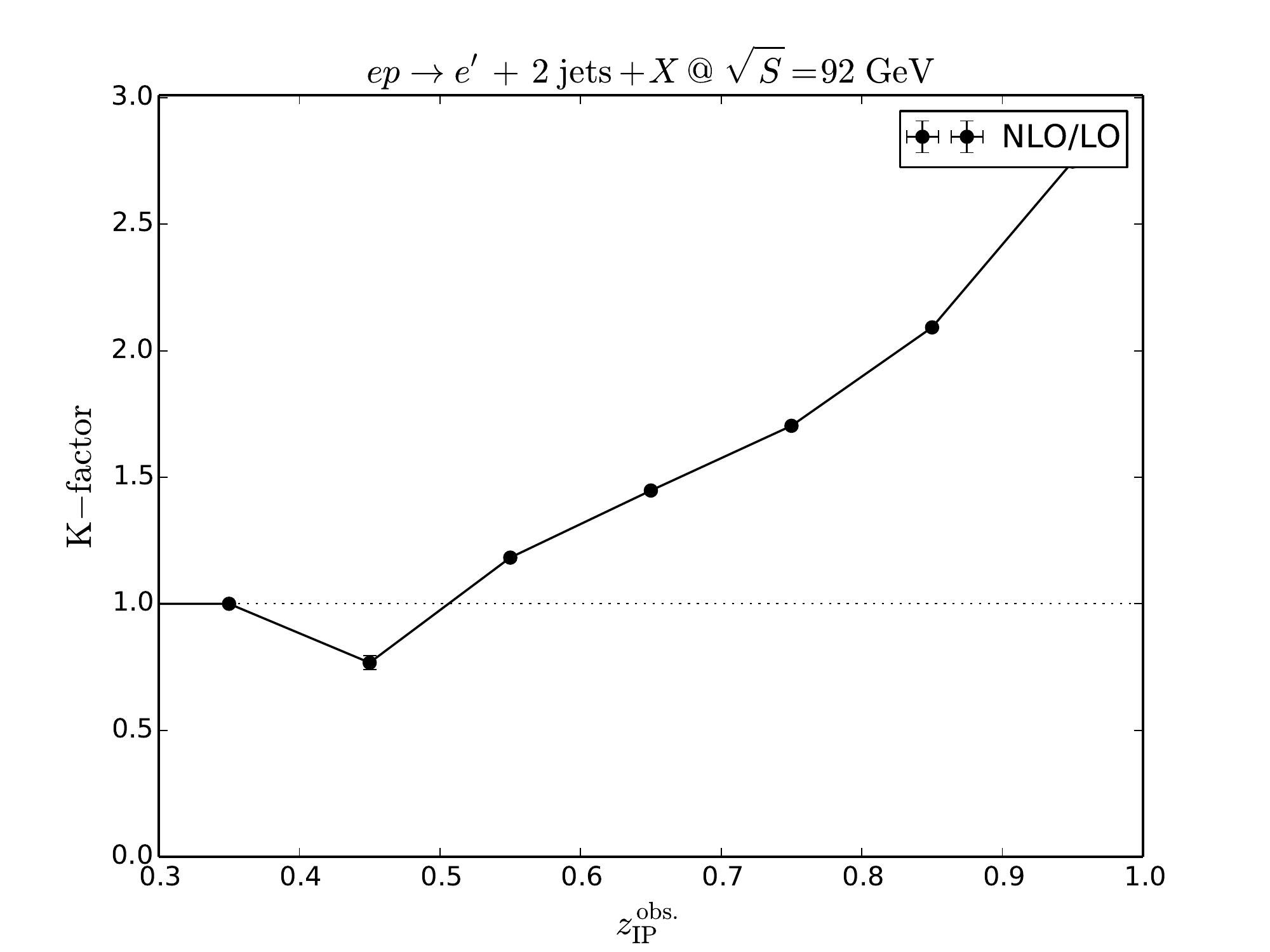,width=0.49\textwidth}
 \caption{$K$-factors for LO cross sections as a function of the jet average
   transverse momentum (top left), jet rapidity difference (bottom
   left) and observed longitudinal momentum fractions in the photon (top right)
   and pomeron (bottom right).}
\label{fig:3}
\end{figure}}
%
sections, as a function of jet average transverse momentum (top left),
jet rapidity difference (bottom left) and observed longitudinal momentum
fractions in the photon (top right) and pomeron (bottom right). At these
low scales, the NLO corrections amount to about a factor of two and are thus
sizable. At the kinematic edges, i.e.\ for large rapidity differences $\Delta
\eta$ or values of $z_{\p}^{\rm obs.}$, they become even larger.
For inclusive jet photoproduction at HERA, the corrections at approximate
next-to-next-to-leading order (aNNLO) increase the cross section for $p_T=20$
GeV by another 12\%, improving the
description of the considered ZEUS data \cite{Abramowicz:2012jz}. However, at
the same time the scale uncertainty is considerably reduced
\cite{Klasen:2013cba}. This demonstrates that the perturbative expansion
remains reliable despite seemingly large $K$-factors at NLO.

\subsection{Evolution of gluon in pomeron contribution}

Although the range in $p_T$ is limited at $\sqrt{S}=92$ GeV, it would be
nice to observe the evolution of the diffractive PDFs in the pomeron with
the energy scale, set here by the average jet transverse momentum. We
therefore compare in Fig.\ \ref{fig:4} the fractional contribution of%
%
{\begin{figure}\centering
 \epsfig{file=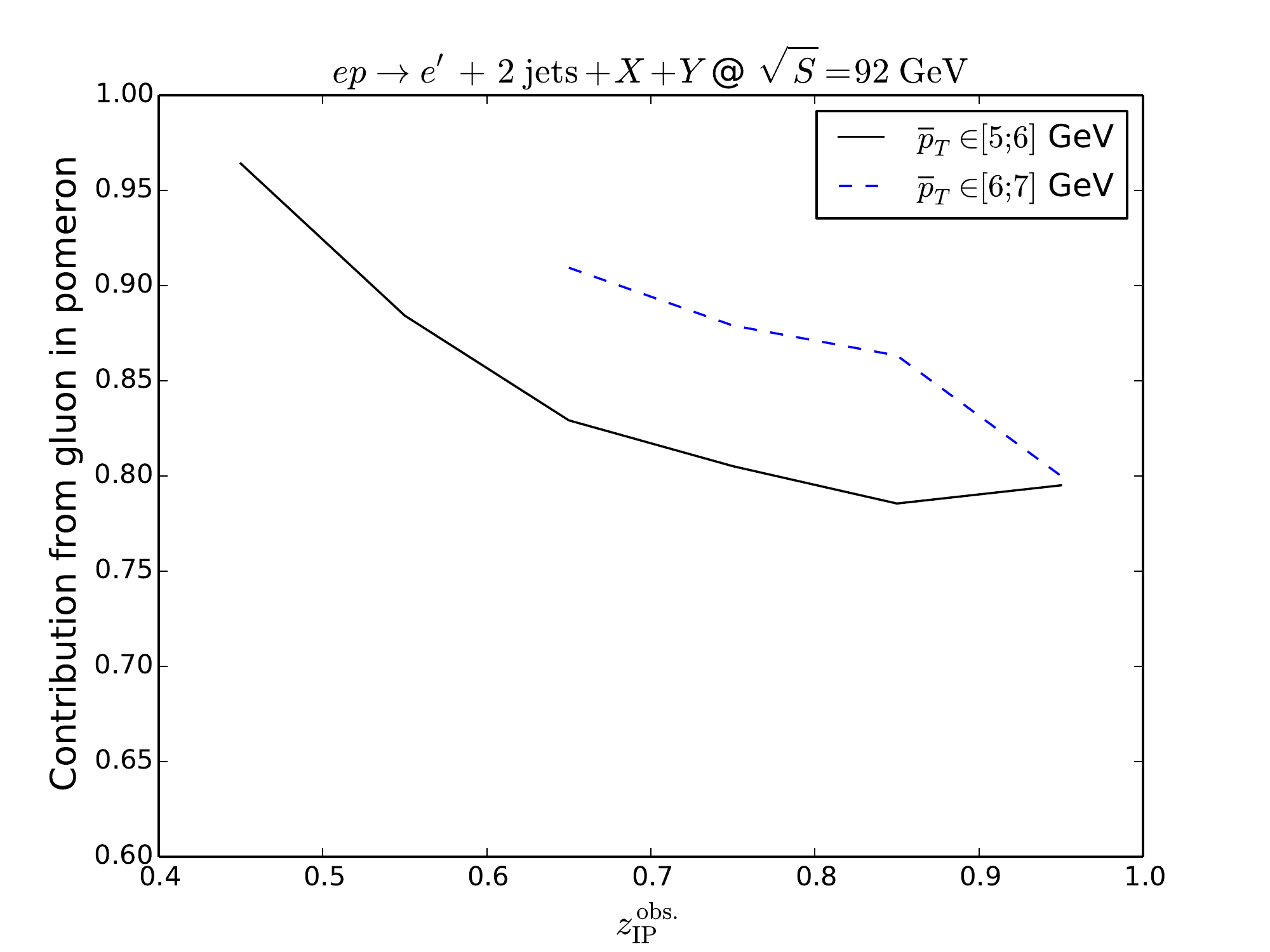,width=0.94\textwidth}
 \caption{Evolution of the contribution from gluons in the pomeron with
   the energy scale, set by the jet average transverse momentum, as a function
   of the longitudinal momentum fraction in the pomeron.}
\label{fig:4}
\end{figure}}
%
the gluon in the pomeron in the lowest $\bar{p}_T$ bin from 5 to 6 GeV to
the one in the next-highest bin from 6 to 7 GeV. Clearly, this restricts the
range in $z_{\p}^{\rm obs.}$ from above 0.4 to even larger values above 0.6.
It also increases the gluon contribution in the pomeron at these large
momentum fractions. This corresponds well with the scale evolution and
relative increase of the rather constant gluon vs.\ the falling quark singlet
density at large $z_{\p}^{\rm obs.}$, as shown in Fig.\ 11 of Ref.\
\cite{Aktas:2006hy}.

\subsection{Dependence on diffractive PDFs}

More important than the evolution of the diffractive PDFs, which should in
principle be predictable from perturbative QCD, is the $z_{\p}$ dependence
itself, which must be determined from experimental data and which is therefore,
despite the considerable progress at HERA, still subject to large
uncertainties. In Fig.\ \ref{fig:5} we therefore compare our NLO QCD%
%
{\begin{figure}\centering
 \epsfig{file=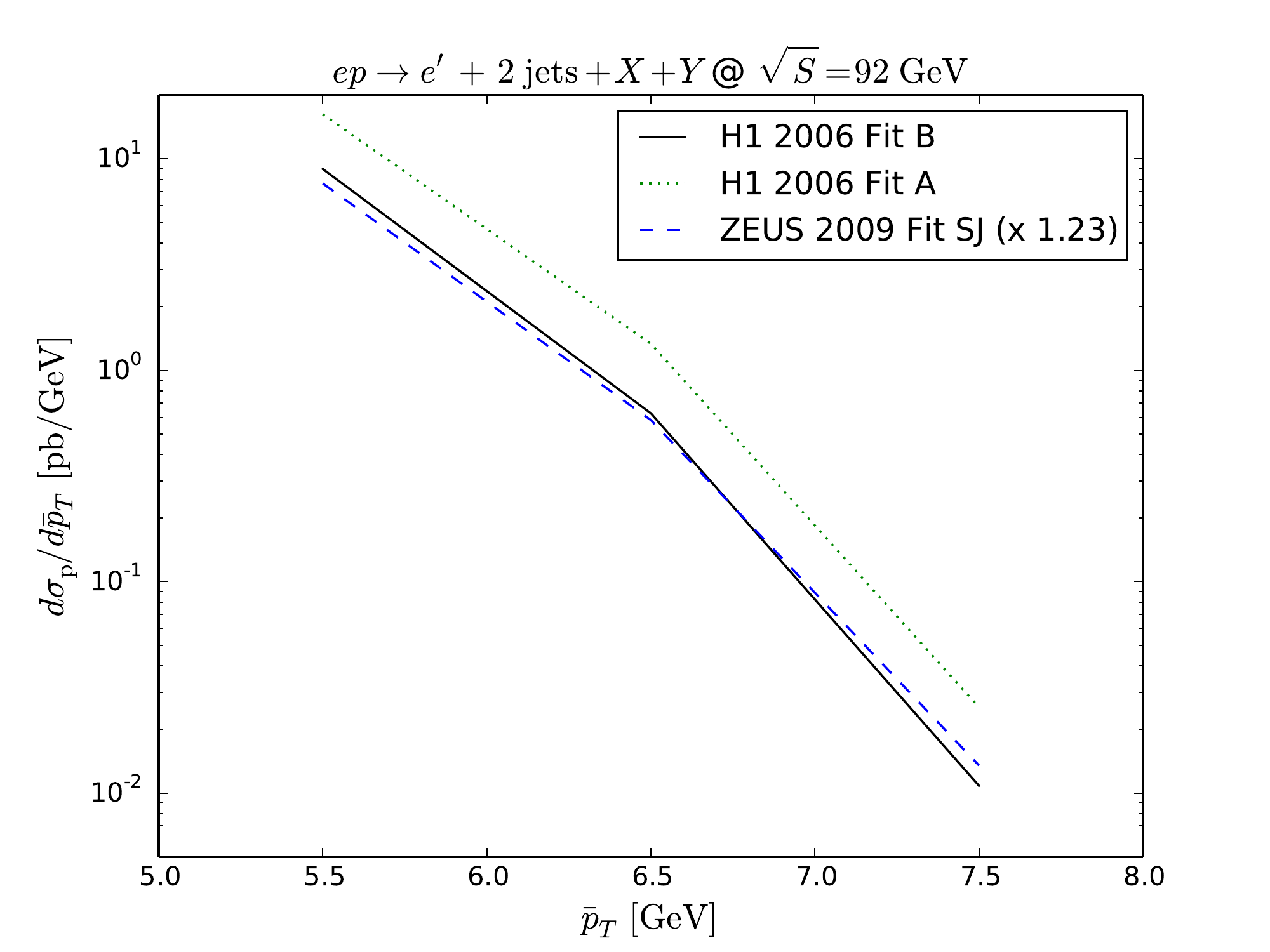,width=0.49\textwidth}
 \epsfig{file=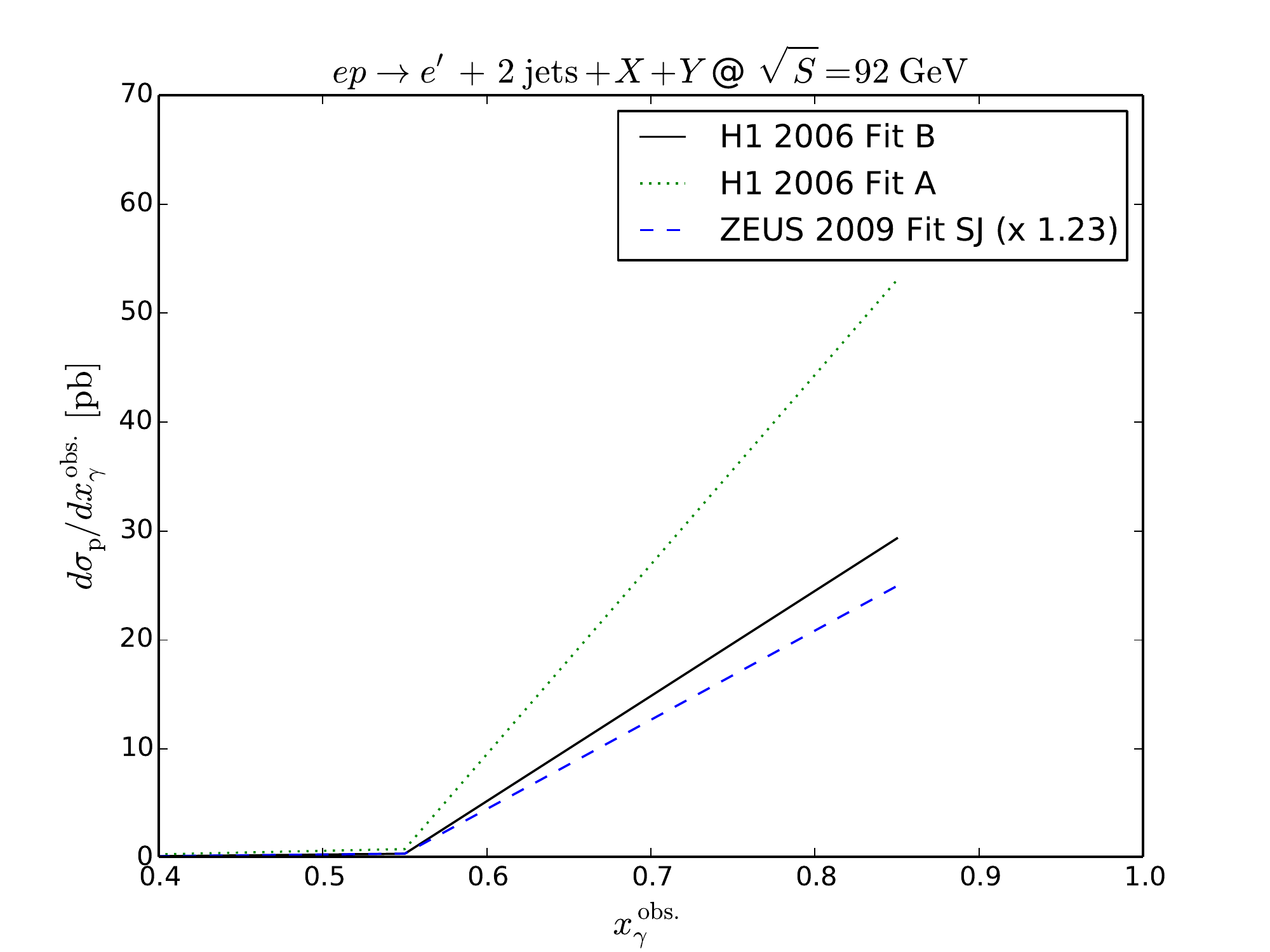,width=0.49\textwidth}
 \epsfig{file=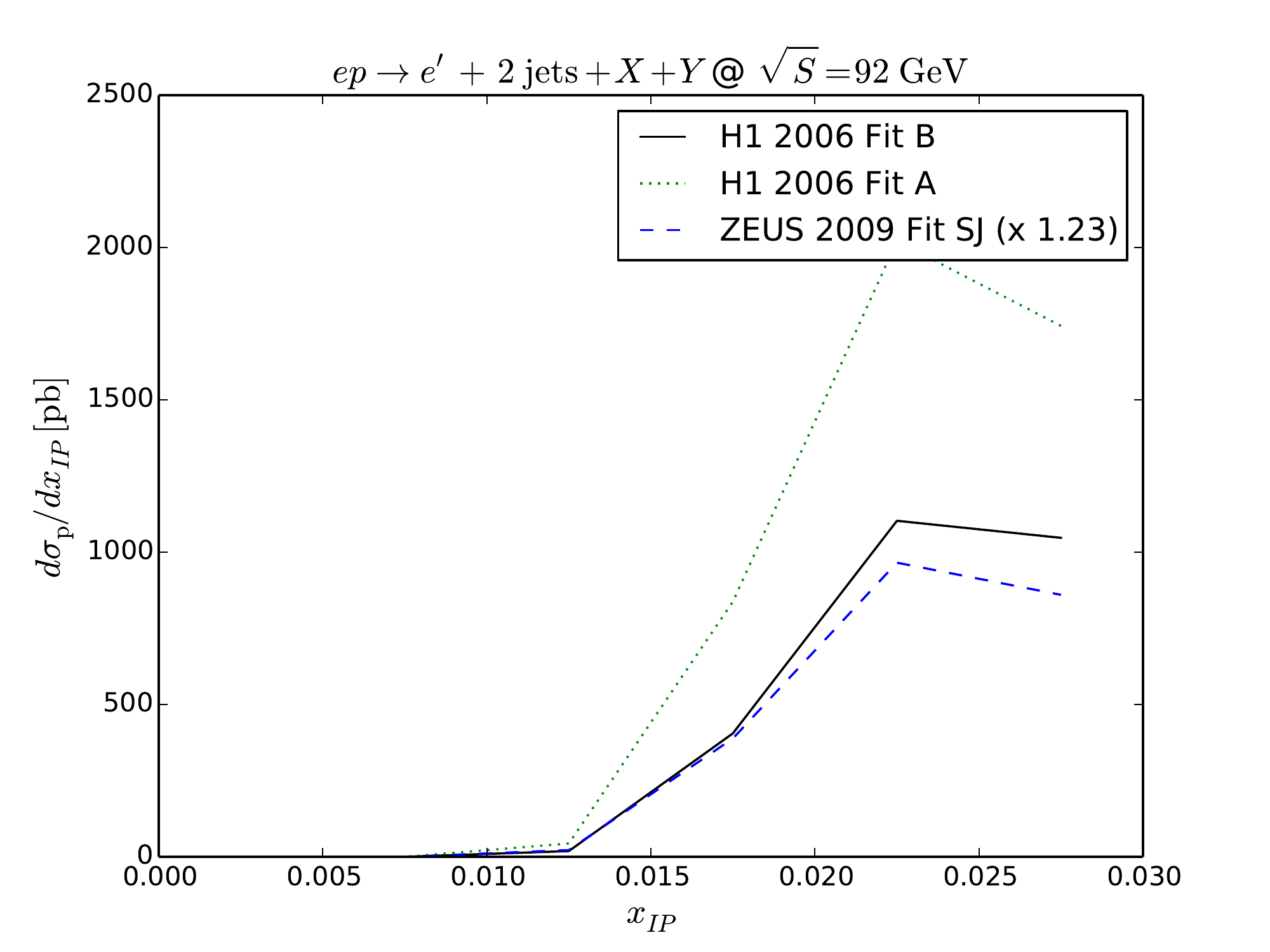,width=0.49\textwidth}
 \epsfig{file=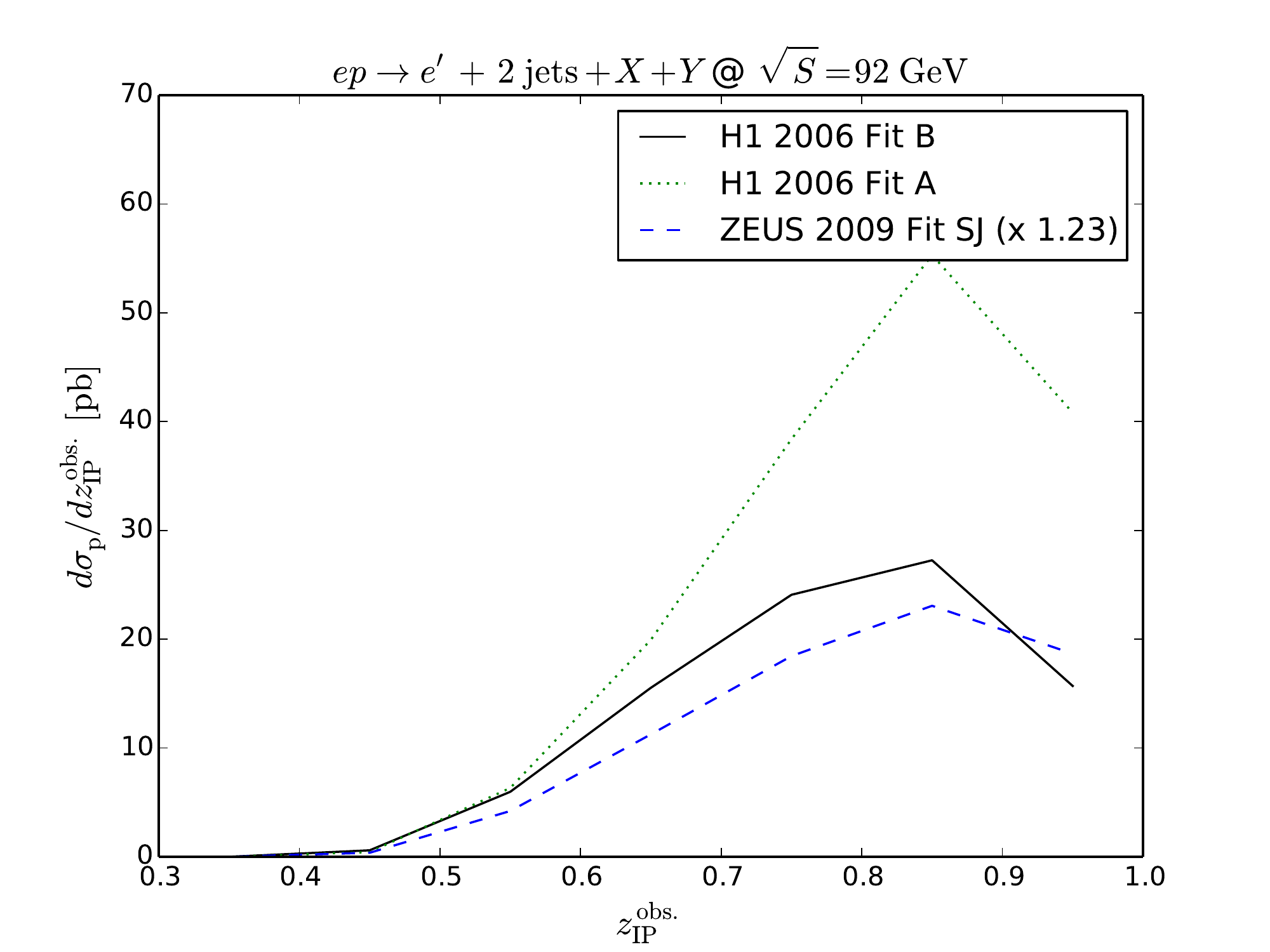,width=0.49\textwidth}
 \caption{Same as Fig.\ \ref{fig:2}, but comparing our NLO QCD predictions
   using three different sets of diffractive PDFs: H1 2006 Fit B (full black),
   Fit A (dotted green), and ZEUS 2009 Fit SJ (dashed blue curves). Since the
   latter have been determined from leading protons, i.e.\ without dissociation
 contributions, they must be rescaled by a factor of 1.23.}
\label{fig:5}
\end{figure}}
%
predictions for the EIC using three different fits of the pomeron PDFs to
diffractive DIS at HERA: our standard prediction with the frequently used
H1 2006 Fit B (full black), the accompanying Fit A (dotted green)
\cite{Aktas:2006hy}, and ZEUS 2009 Fit SJ (dashed blue curves)
\cite{Chekanov:2009aa}. Since the
latter has been obtained from leading protons, i.e.\ without dissociation
contributions, the corresponding cross sections must be and have been
multiplied by a factor of 1.23. The main differences between H1 2006 Fits A
and B are the starting scales $Q_0^2=1.75$ GeV$^2$ and 2.5 GeV$^2$,
respectively, and the
gluon parametrization at large $z_{\p}$, which is singular in Fit A and --
up to the small-$z_{\p}$ exponential term -- constant in Fit B. More precisely,
both the gluon and singlet quark densities are parametrized at the starting
scale as
\beq
 z_{\p}f_i(z_{\p},Q_0^2)=A_iz_{\p}^{B_i}(1-z_{\p})^{C_i},\quad i=g,q,
\eeq
where $C_g=-0.95\pm0.20$ in Fit A and $C_g$ is fixed to 0 in Fit B. Attempts
have subsequently been made to reduce this uncertainty by adding to the
inclusive data also jet production data as in H1 2007 Fit Jets (not used)
\cite{Aktas:2007bv} and ZEUS 2009 Fit SJ \cite{Chekanov:2009aa}. The former
uses again $Q_0^2=2.5$ GeV$^2$ and results in $C_g=0.91\pm0.18$, the latter
uses $Q_0^2=1.8$ GeV$^2$ and results in the smallest uncertainty on
$C_g=-0.725\pm0.082$. Note, however, that $C_g$ is intimately linked to the
other parameters in the gluon and quark singlet fits, including the pomeron
flux factor, so that they cannot be directly compared. What one can observe
from Fig.\ \ref{fig:5} is that the predictions based on H1 2006 Fit A rise
indeed much more steeply as $z_{\p}\to1$ and that H1 2006 Fit B, 2007 Fit Jets
(not shown) and ZEUS 2009 Fit SJ give comparable results.

\subsection{Range in $x_{\p}$ and reggeon contribution}

The observations of the rather limited range in transverse momentum and the
overwhelming importance of the direct photon contribution, that leaves little
hope for resolving the question of factorization breaking, motivate us to
consider also a larger range in $x_{\p}$. In Fig.\ \ref{fig:6} we therefore%
%
{\begin{figure}\centering
 \epsfig{file=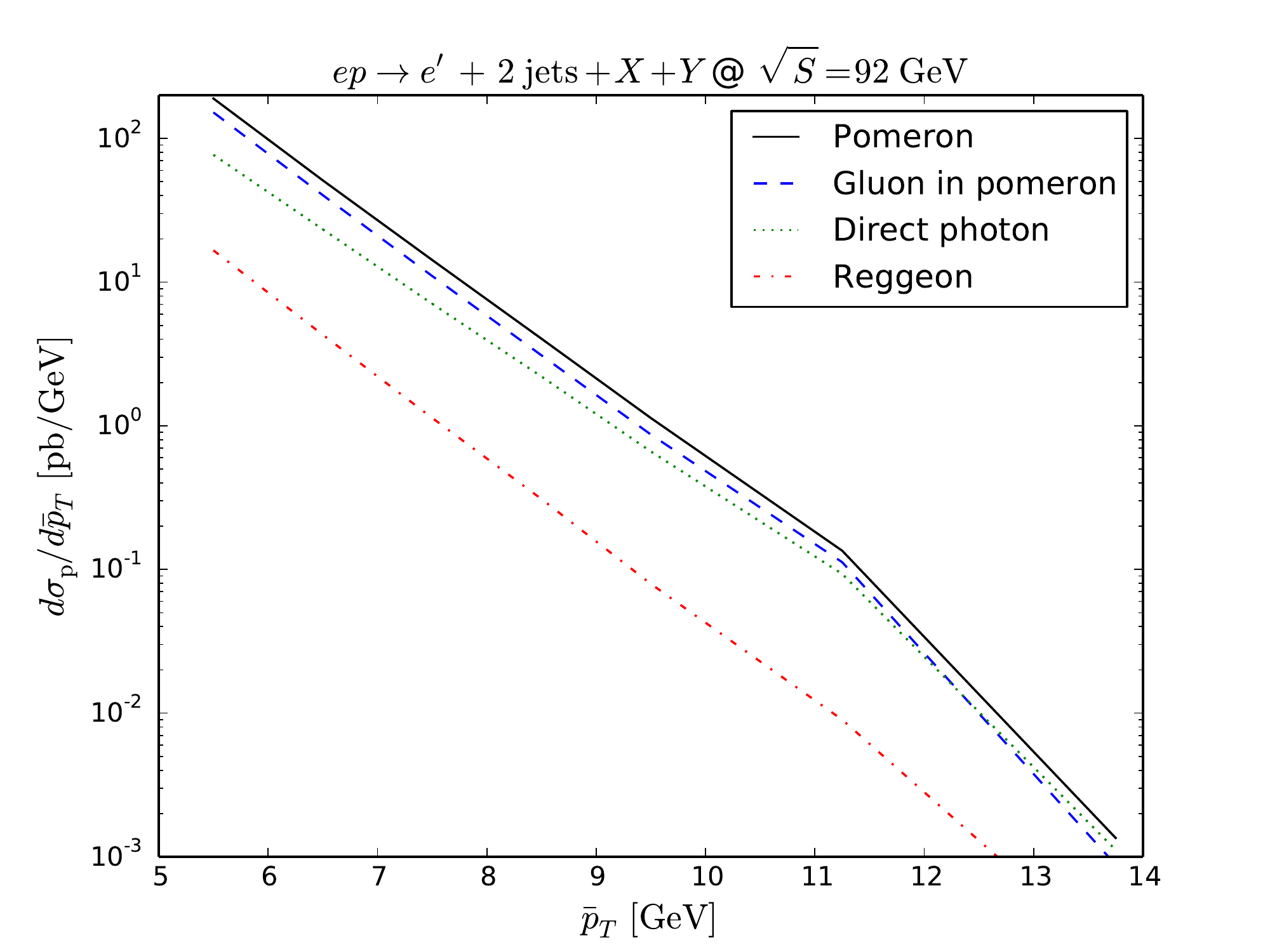,width=0.49\textwidth}
 \epsfig{file=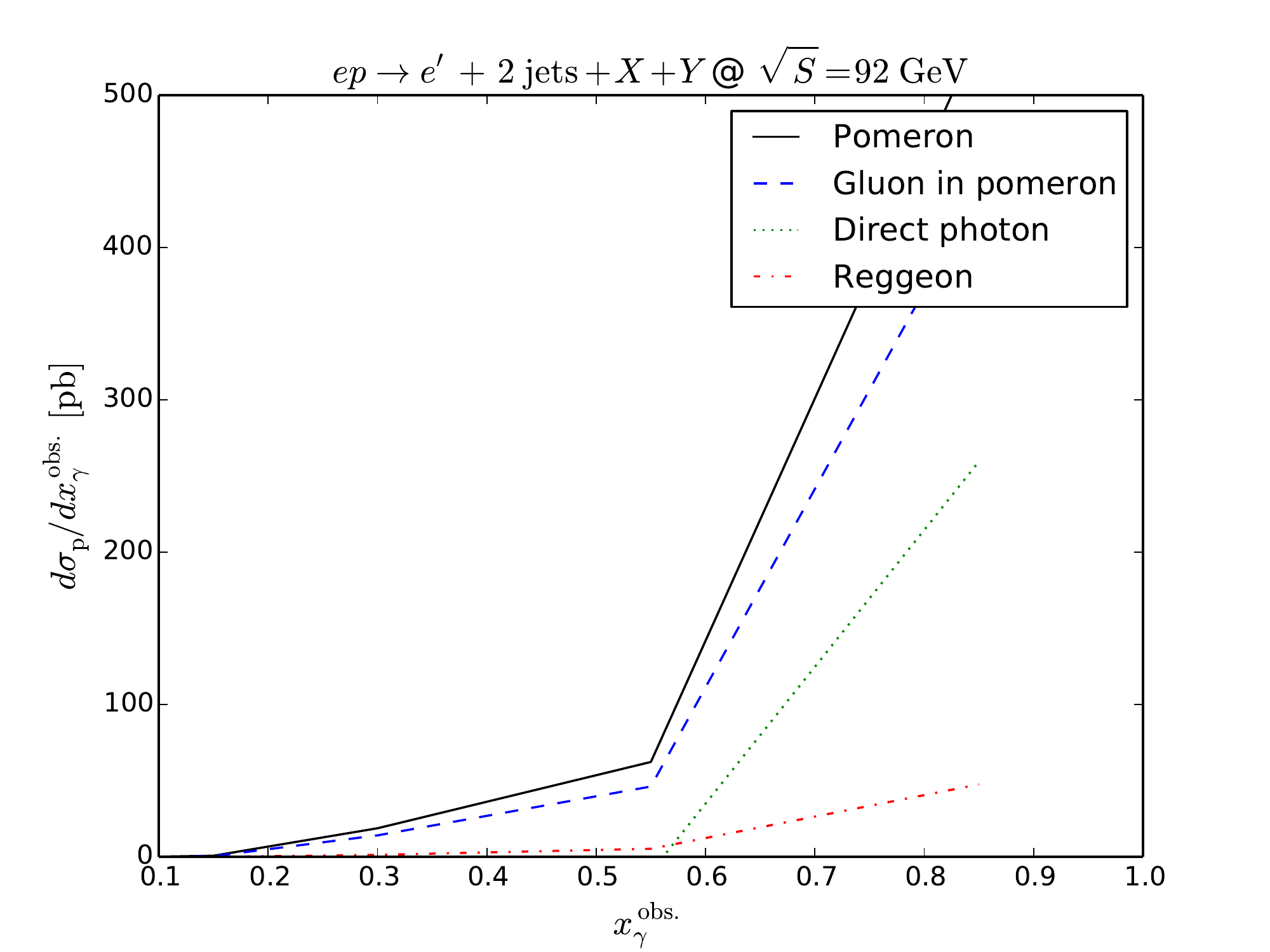,width=0.49\textwidth}
 \epsfig{file=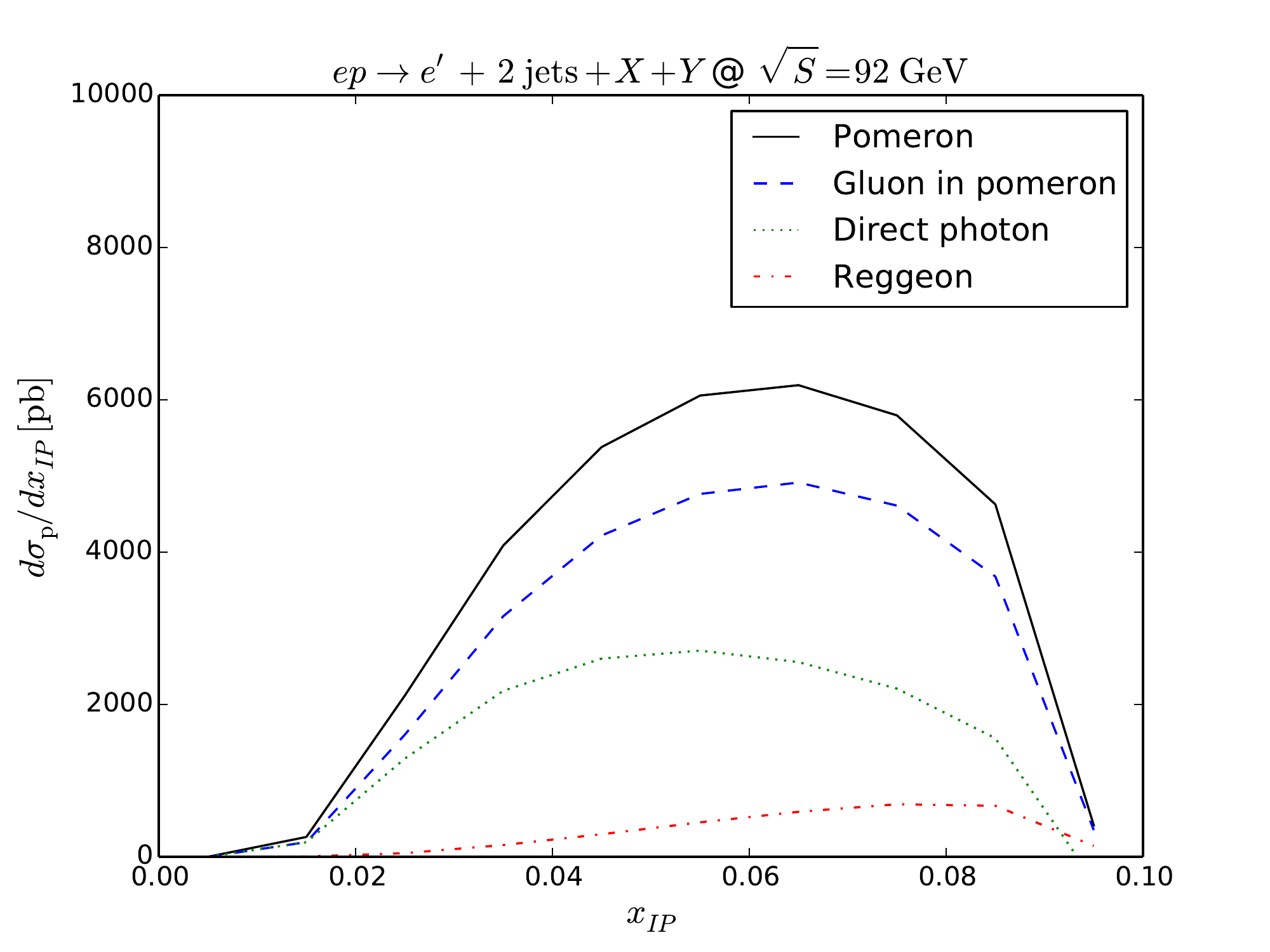,width=0.49\textwidth}
 \epsfig{file=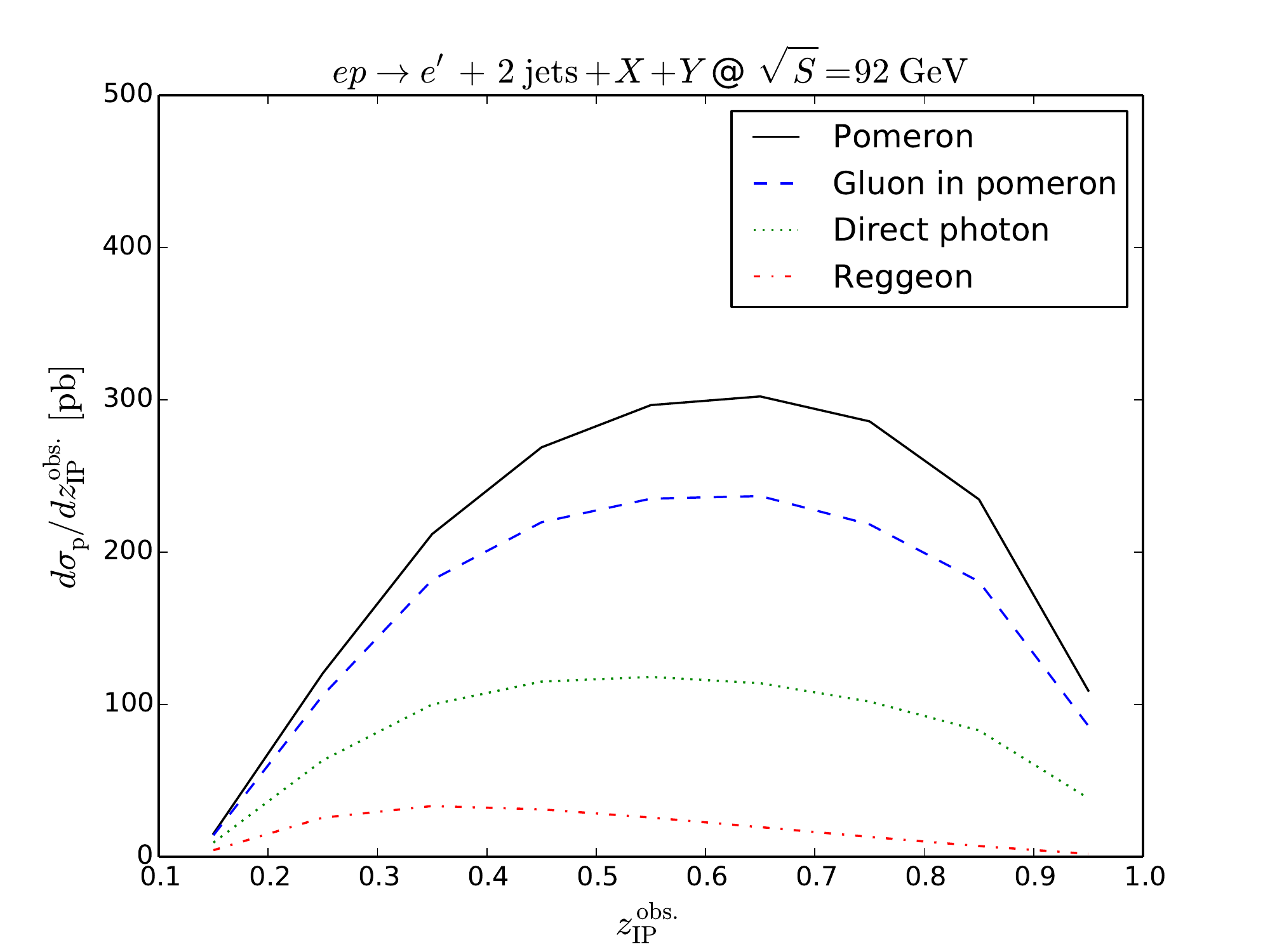,width=0.49\textwidth}
 \caption{Same as Fig.\ \ref{fig:2}, but now with an extended range in $x_{\p}
   <0.1$. In addition, also the contribution from the subleading reggeon
 is shown (dot-dashed red curves).}
\label{fig:6}
\end{figure}}
%
extend it from $x_{\p}<0.03$ to $x_{\p}<0.10$. This immediately enlargens the
reach
in $\bar{p}_T$ from 8 to 14 GeV (top left) and the momentum fraction in the
photon from 0.5 down to 0.1 (top right), so that now also resolved photons
contribute substantially. Furthermore, the PDFs in the pomeron can now be
probed in the entire range of $z_{\p}$ from 0.1 to 1 (bottom right).
The increase also seems to be sufficiently large, as the distribution
in $x_{\p}$ is no longer peaked at the cut, but around 0.06 (bottom
left). This is important
since the contribution from the subleading reggeon
trajectory increases from less than about 2\% at $x_{\p}\leq0.03$ to
$10-35\%$ at $x_{\p}\geq 0.06-0.10$. In fact, to obtain a good description of
the HERA diffractive DIS data, H1 and ZEUS include an additional sub-leading
exchange ($\reg$), which has a lower trajectory intercept than the pomeron and
which contributes significantly only at large $\xpom$ and low $z_{\p}$. This
contribution is assumed to factorize in the same way as the pomeron term, such
that the diffractive PDFs take the form
\begin{equation}
f_{i/p}^D(x,Q^2,\xpom,t) = f_{\pom/p}(\xpom,t) \cdot f_{i/\p} (z_{\p},Q^2) \ + \
n_\reg \cdot f_{\reg/p}(\xpom,t) \cdot f_{i/\reg} (z_{\p},Q^2) \ .
\label{reggefac2}
\end{equation}
The flux factor $f_{\reg/p}$ takes the form of Eq.\ (\ref{eq:fluxfac}), 
normalised via a parameter $A_\reg$ in the same manner as for the pomeron
contribution and with fixed parameters $\alpha_\reg(0)$, $\alpha_\reg^\prime$%
%
{\begin{figure}\centering
 \epsfig{file=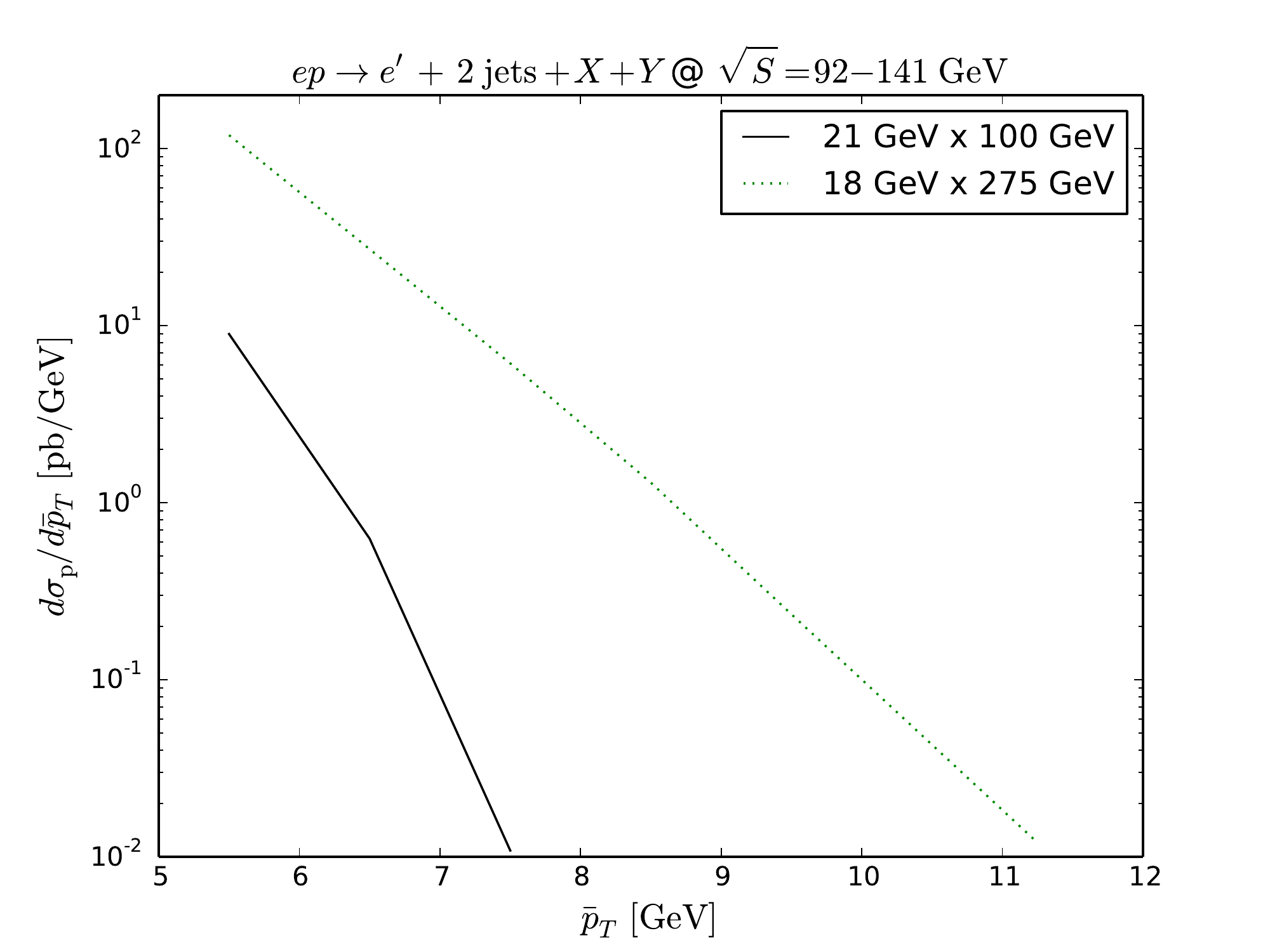,width=0.49\textwidth}
 \epsfig{file=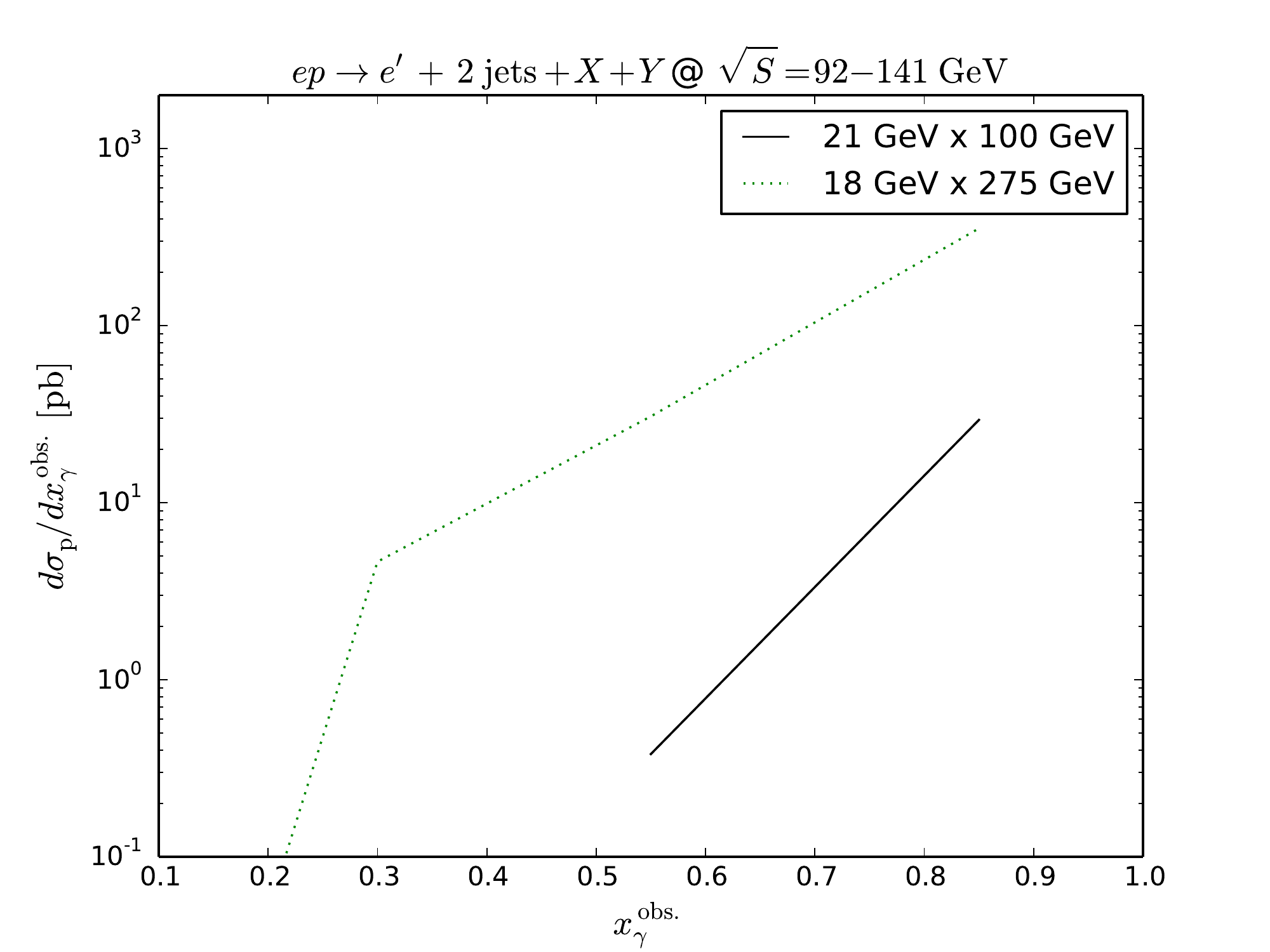,width=0.49\textwidth}
 \epsfig{file=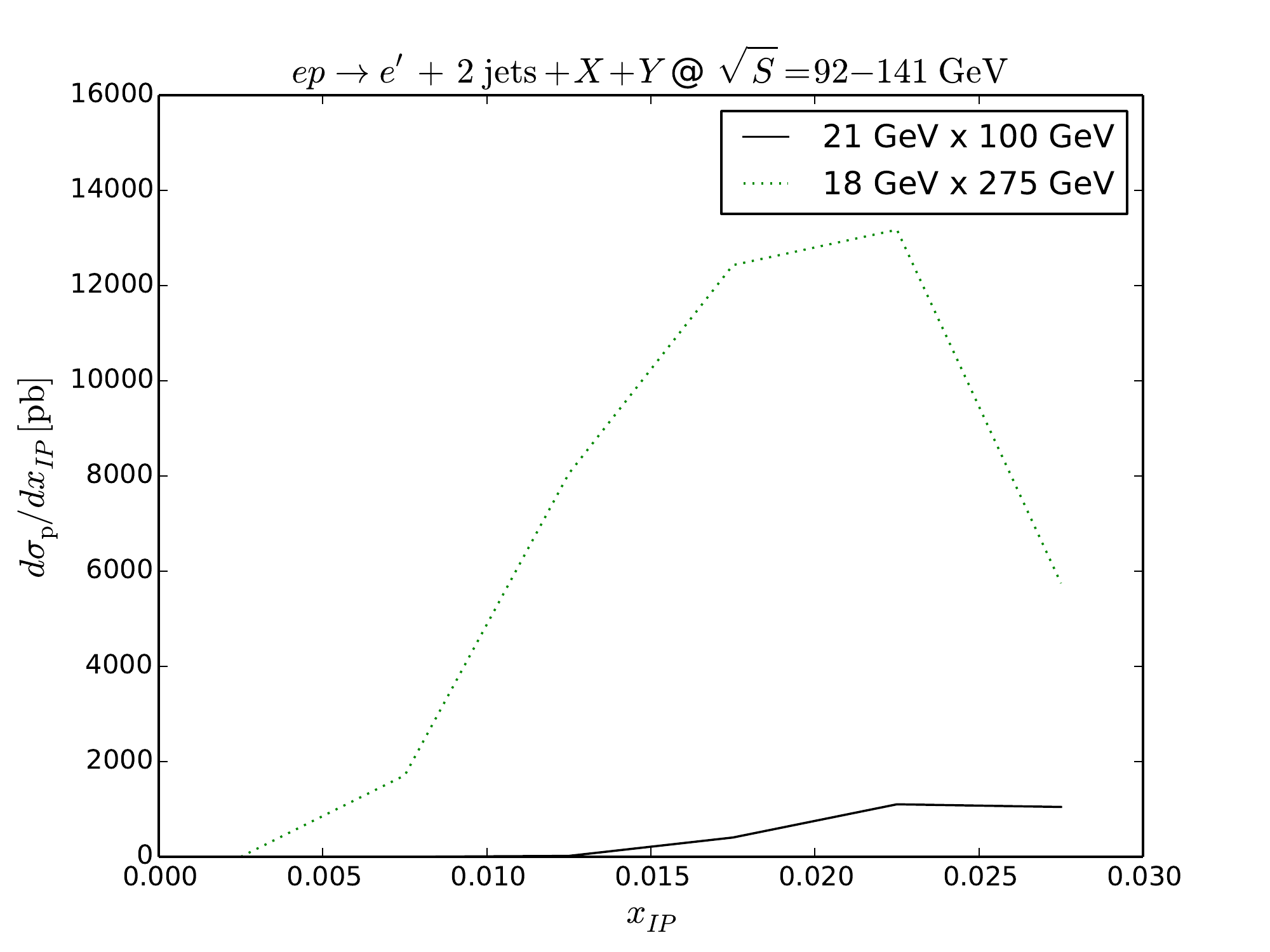,width=0.49\textwidth}
 \epsfig{file=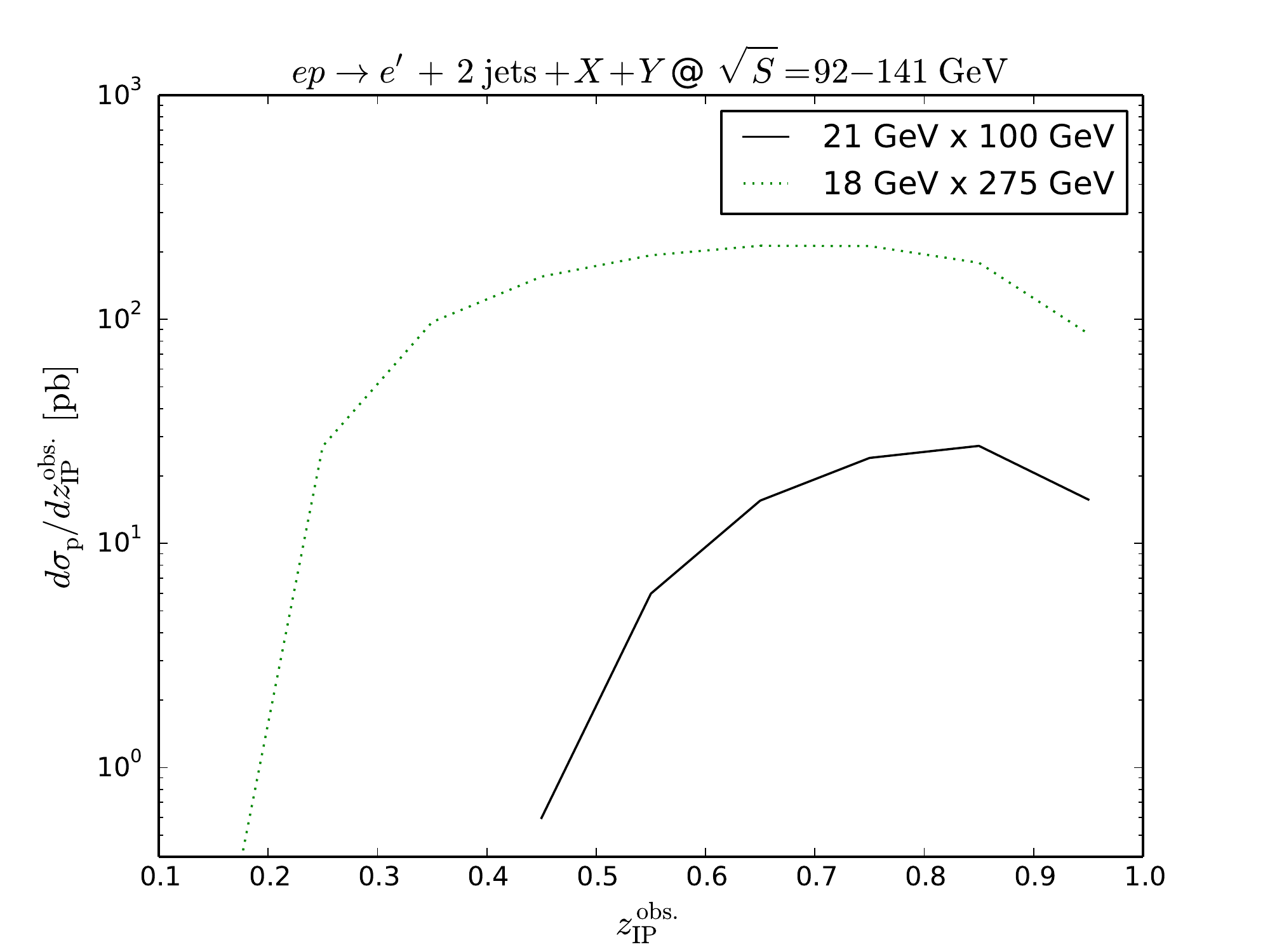,width=0.49\textwidth}
 \caption{Differential diffractive dijet photoproduction cross sections for
   collisions of 18 GeV electrons with 275 GeV protons at a CMS energy of
   $\sqrt{S}=141$ GeV (dotted green) compared to our default cross sections
   for 21 GeV electrons with 100 GeV protons at a CMS of 92 GeV (full black
   curves).}
\label{fig:7}
\end{figure}}
%
and $B_\reg$ obtained from other H1 and ZEUS measurements \cite{Aktas:2006hy,%
  Aktas:2007bv,Chekanov:2009aa}. The parton densities $f_{i/\reg}$ of the
sub-leading exchange are taken from fits to pion structure function data. We
choose the GRV NLO parametrization \cite{Gluck:1991ey}. Other pion PDFs give
similar results \cite{Aktas:2006hy}.

\subsection{EIC beam energy dependence}

Like RHIC, the EIC will accelerate bare charged protons to higher energies per
nucleon than it will accelerate nuclei that comprise also neutrons, i.e.\ not
only to 100 GeV, but even to $E_p=275$ GeV. At the same time, the most recent plans envisage an electron beam of energy $E_e=18$ GeV rather than 21 GeV
\cite{Skaritka:2018oxj}. In Fig.\ \ref{fig:7}, we therefore repeat our studies
for this accelerator design and compare the reach in the different distributions with our default predictions.
While a different hadron beam energy will make the extraction of nuclear
effects from comparisons of the bare proton baseline with heavy nuclei more
difficult, it increases the reach in the kinematic variables relevant for
diffraction studies on protons alone. A first example is the reach in average
transverse momentum $\bar{p}_T$, which is extended from 8 to 12 GeV (top left).
Interestingly, an increase in the cut on $x_{\p}$ to 0.1 had a larger effect,
extending the reach to 14 GeV (cf.\ Fig.\ \ref{fig:6}). A combination of both
will therefore lead to an even larger reach. The
$x_{\p}$-distribution itself (bottom left) is now broader and its maximum near
the cut at 0.0225 less sharp. Similarly to what we observed in the previous
section with a larger $x_{\p}$ range, the
distributions in longitudinal momentum fraction in the photon (top right) and
pomeron (bottom right) also span now larger regions, here from 0.2
(rather than 0.1) to 1. In addition, the corresponding
differential cross sections are now larger by one to two orders of magnitude,
leading to increased statistics and precision in the corresponding
measurements.

\subsection{Factorization breaking}
\label{sunsec:fact_break}

Factorization breaking in diffractive dijet photoproduction is a result of soft inelastic photon interactions with the proton, which populate and thus partially destroy the final-state rapidity gap. This effect is usually described in the literature by a rapidity gap survival factor $S^2 \leq 1$. Since the magnitude of $S^2$ decreases with an increase of the interaction strength between the probe and the target, the pattern of the factorization breaking can be related to various components of the photon 
\cite{Kaidalov:2009fp}. In the laboratory reference frame, the high-energy photon interacts with hadronic targets by fluctuating into various configurations (components) interacting with the target with different cross sections. These fluctuations contain both weakly-interacting (the so-called point-like) components and the components interacting with large cross sections, which are of the order of the vector meson-proton cross sections.
This general space-time picture of photon-hadron interactions at high energies 
is usually realized in the framework of such approaches as the vector meson dominance (VMD) model and its generalizations~\cite{Bauer:1977iq} or the color dipole model~\cite{Nemchik:1996pp,Kowalski:2006hc}. 
It is also used in the language of collinear factorization, where 
the photon structure function and parton distribution functions (PDFs) are given by a sum of the resolved-photon contribution
corresponding to the VMD part of the photon wave function and the point-like (inhomogeneous) term
originating from the $\gamma \to q \bar{q}$ splitting, see, e.g., Ref.~\cite{Gluck:1991jc}.
Note that the direct-photon contribution to diffractive dijet photoproduction corresponds to
the configurations interacting with very small cross sections of the order of $1/p_T^2$, which preserves factorization. 

The mechanism of factorization breaking in photoproduction is one of the
important desiderata of diffraction studies at HERA and could be one of the
physics goals of the EIC. The key question is whether factorization still
holds for pointlike photons, similarly to DIS, where it has been proven
\cite{Collins:1997sr}, and is only broken for resolved photons  \cite{Klasen:2005dq}, similarly to hadron-hadron scattering, where it is known to be broken \cite{Affolder:2000vb,Sirunyan:2020ifc,Klasen:2009bi,Khoze:2000wk},
whether it is broken globally to a significant \cite{Aktas:2007hn} or only a small extent \cite{Chekanov:2007rh}, as the H1 and ZEUS data seem to indicate
\cite{Klasen:2008ah}, and whether pointlike photon fluctuations into all quark-antiquark pairs \cite{Klasen:2005dq} or rather those of light quark
flavors only \cite{Guzey:2016awf} rescatter and destroy the rapidity gap.
This last point can be related to the mass scheme employed in the photon structure function, i.e.\ the applicability of dimensional regularization and
the zero-mass variable flavor number scheme (ZM-VFNS) vs.\ the fixed flavor number scheme (FFNS), where the heavy quark mass serves as a regulator.
Unfortunately, current photon structure function data do not yet allow 
to determine the corresponding kinematic ranges. Theory and experience from
proton PDFs tells us that the FFNS should be used when $p_T\leq m_q$, while the ZM-VFNS should be used when $p_T\gg m_q$. In diffractive photoproduction both at HERA and the EIC, we find ourselves in the transition region
$p_T\geq m_q$, so that (in particular charm) mass effects are indeed still relevant and the charm contribution does not seem to be suppressed.

A precondition for an important contribution from the EIC to settle these
questions is a sufficiently large range in direct and resolved photon
contributions, i.e.\ in $x_{\gamma}$. For this reason, we continue to base
our numerical studies here on the accelerator design studied in the last subsection with its higher proton beam energy of $E_p=275$ GeV. To avoid the
reggeon contribution, which could obscure the situation further, we use the lower cut on $x_{\p}<0.03$. In Fig.\ \ref{fig:8} we compare three different%
%
{\begin{figure}\centering
 \epsfig{file=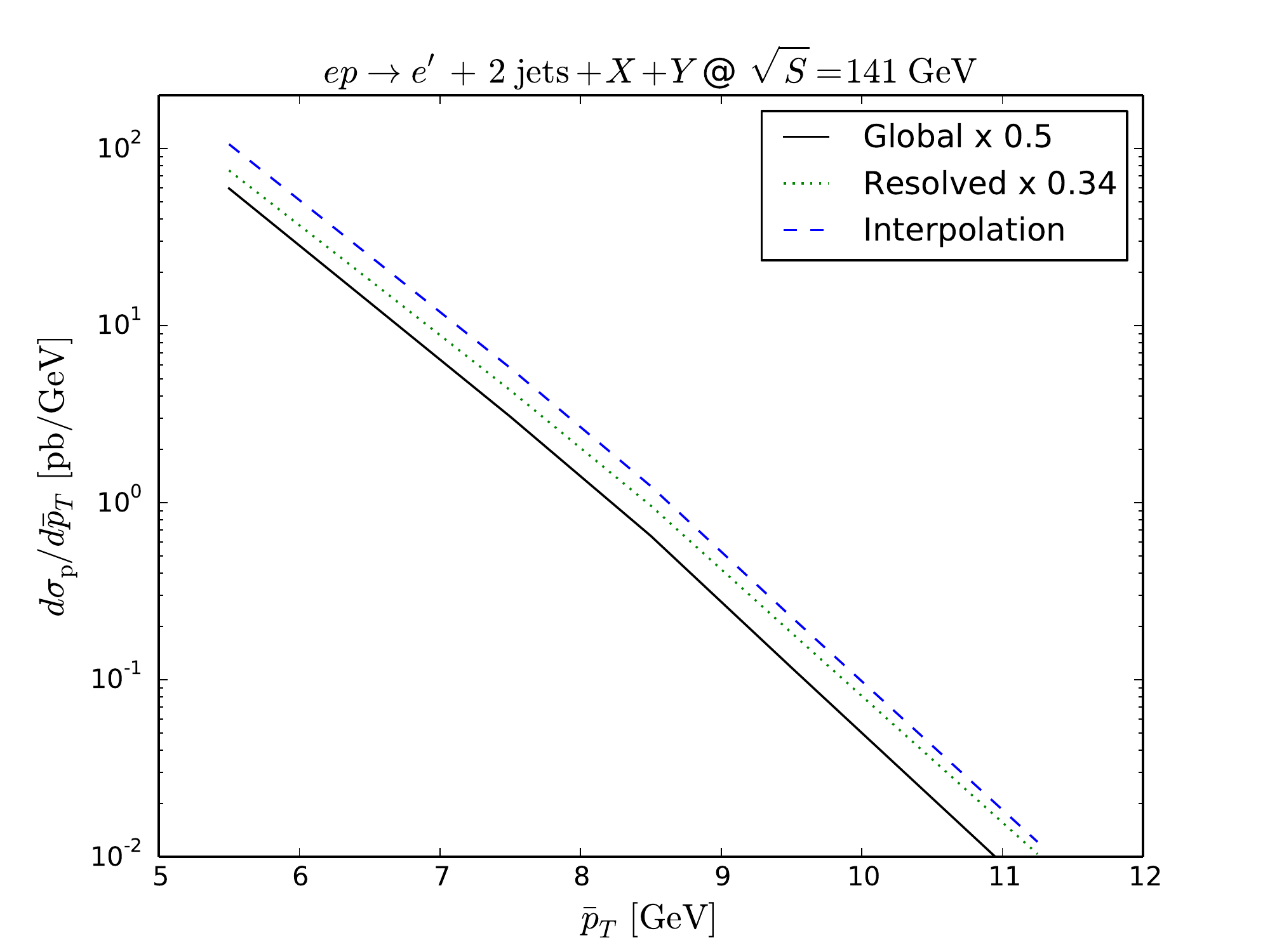,width=0.49\textwidth}
 \epsfig{file=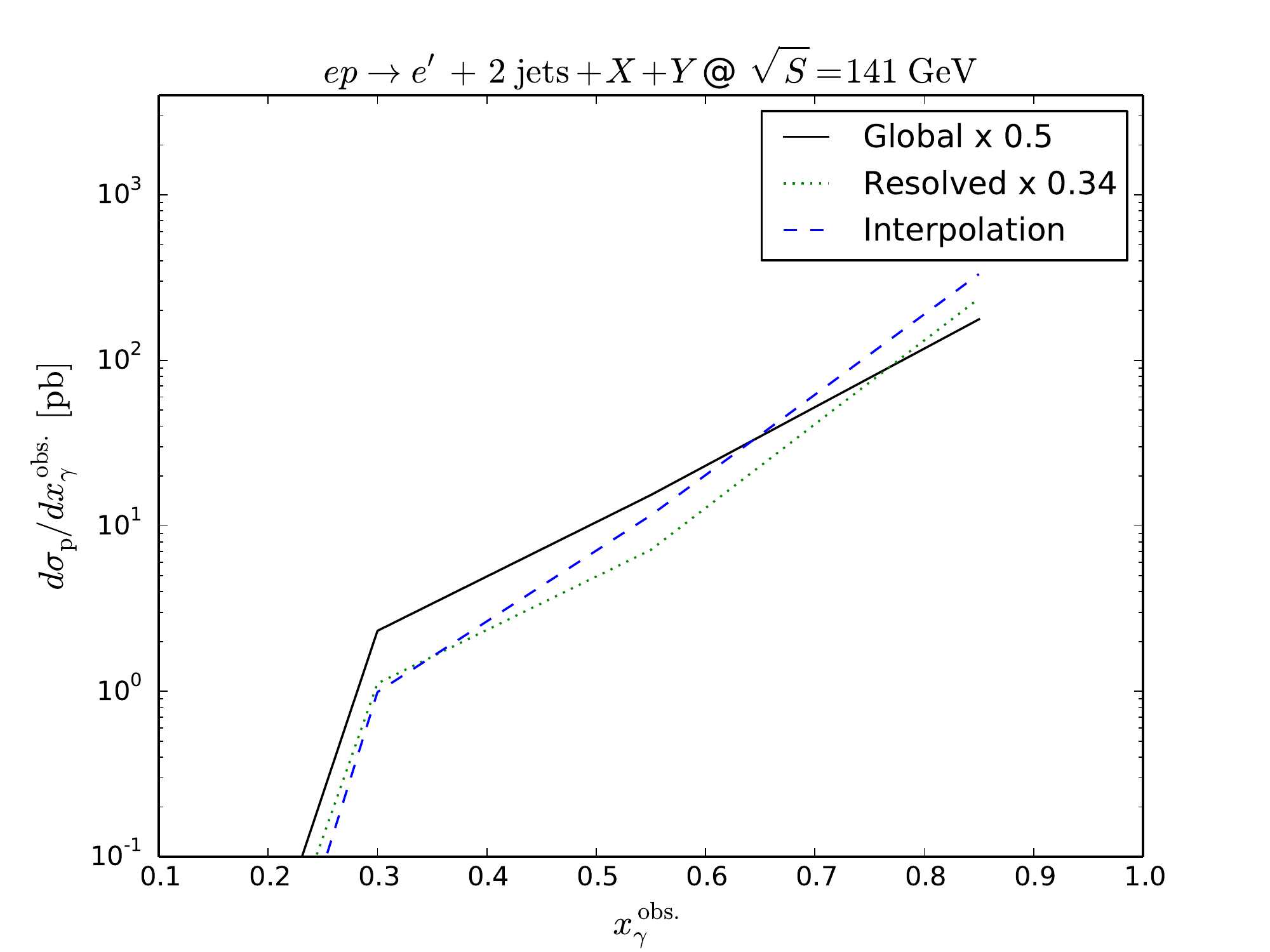,width=0.49\textwidth}
 \epsfig{file=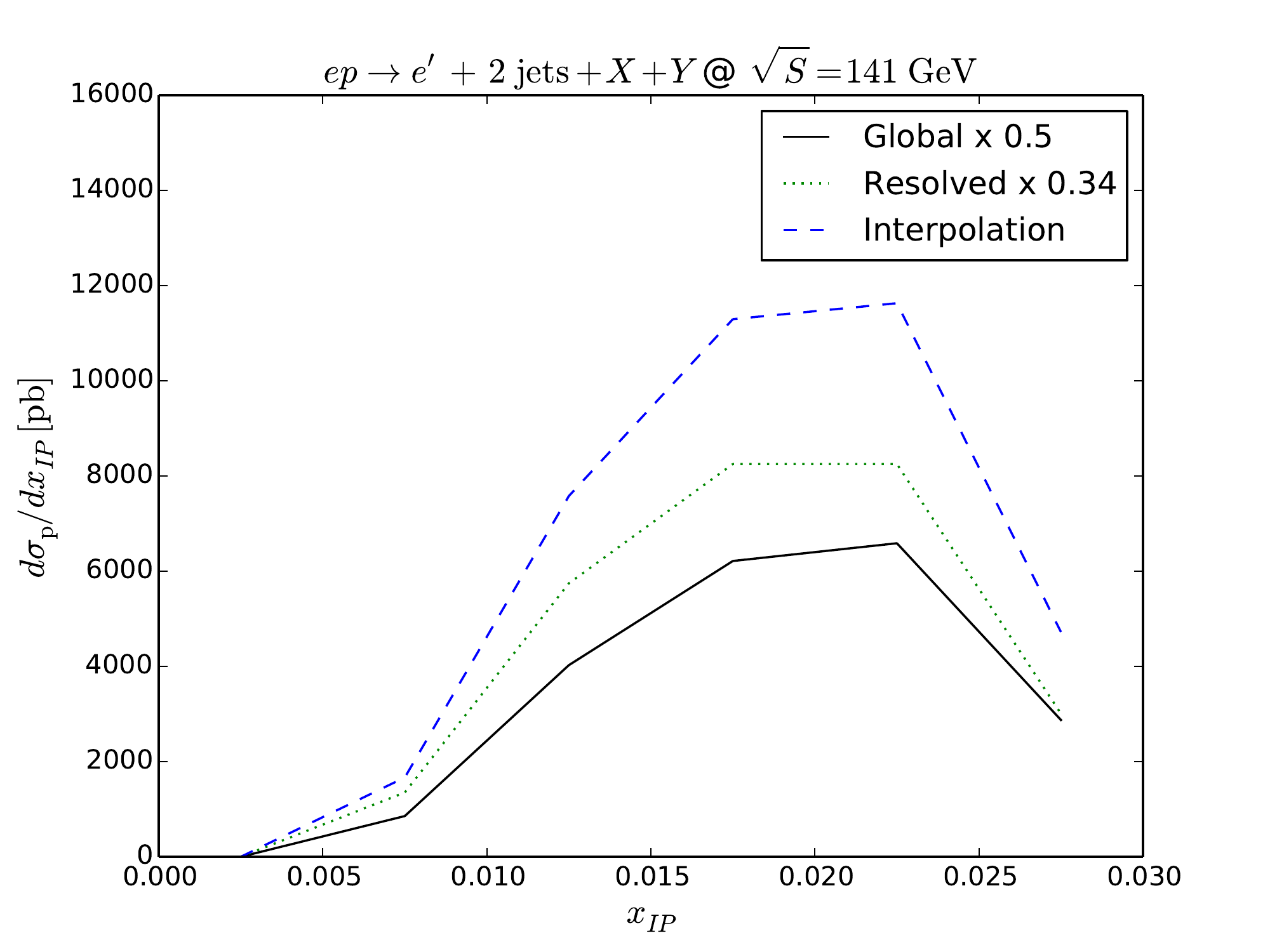,width=0.49\textwidth}
 \epsfig{file=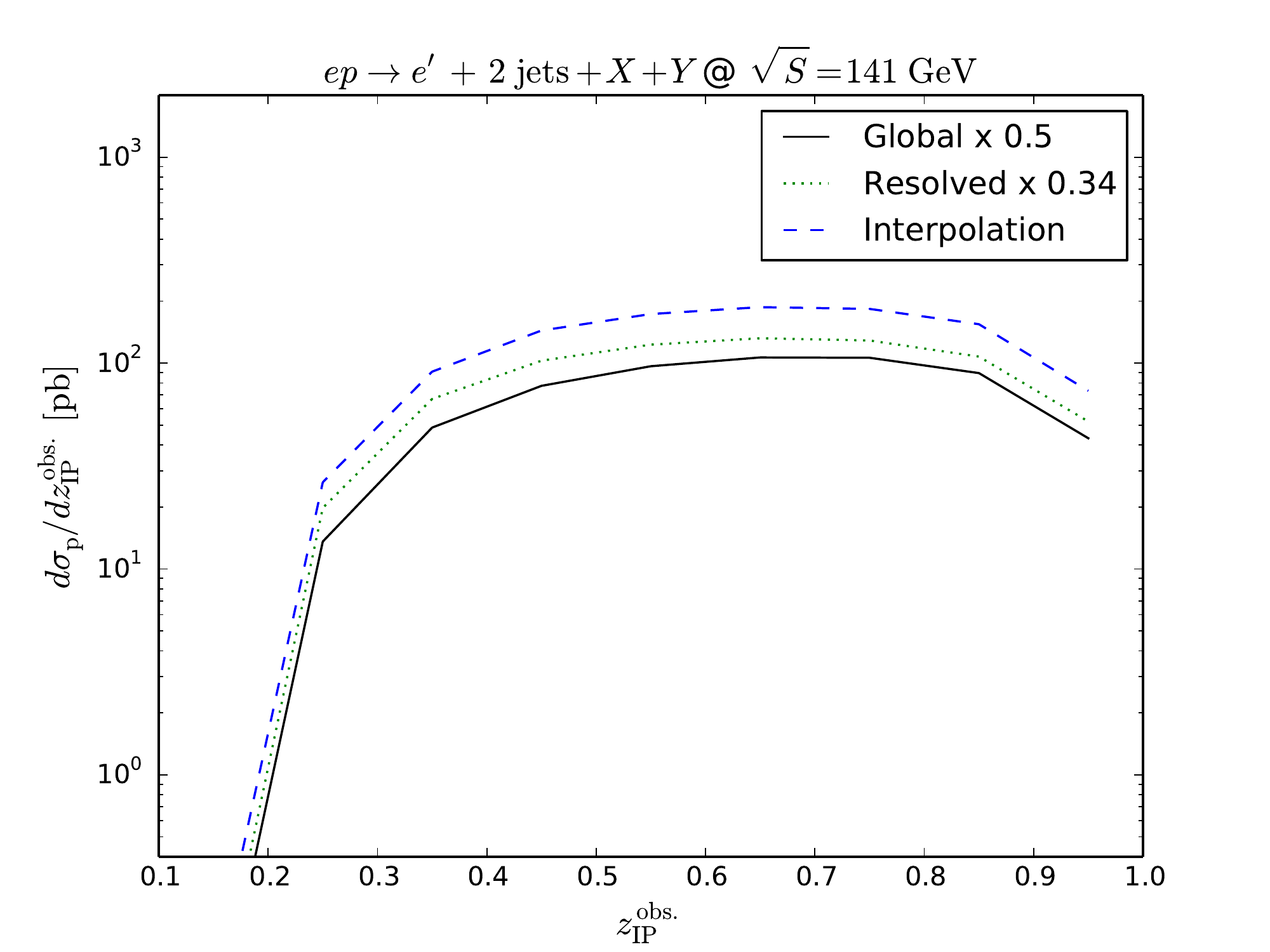,width=0.49\textwidth}
 \caption{Comparison of factorization breaking schemes in diffractive dijet
   photoproduction at the EIC for collisions of electrons with energy 18 GeV and protons of energy 275 GeV. Shown are NLO QCD predictions with a global suppression by a factor of 0.5 (full black), suppression of resolved-photon contributions only by a factor of 0.34 (dotted green) and of an interpolated suppression factor that depends on the type of parton in the photon and $x_{\gamma}$ (see text, dashed blue curves).}
\label{fig:8}
\end{figure}}
%
schemes of factorization breaking,
i.e.\ a global factorization breaking by a factor of 0.5 as determined in the low-$p_T$ measurement of H1 (full black curves) \cite{Aktas:2007hn}, a breaking of the resolved-photon components by a factor of 0.34 as predicted
by the two-channel eikonal model (dotted green curves) \cite{Kaidalov:2003xf}, which is not substantially altered by the finite remainder of collinear quarks and antiquarks \cite{Klasen:2005dq}, and a scheme that interpolates the suppression for photon components of different size as a function of $x_{\gamma}$ \cite{Guzey:2016awf}. In this last scheme, one expects $S^2 \approx 0.34$ for the hadron-like component of the photon at small $x_{\gamma}$, and $S^2 \approx 0.53-0.75$ for the gluon and quark contributions
at large $x_{\gamma}$ corresponding to small, but non-negligible factorization breaking due to the point-like component of the resolved photon \cite{Kaidalov:2009fp,Klasen:2010vk}. Therefore, we interpolate the effect of factorization breaking with
\begin{equation}
S_{i}^2(x_{\gamma}) \to  \left\{\begin{array}{ll}
1 \,, & i=c \,, \\
A_q\, x_{\gamma} +0.34 \,, \quad & i=u,d,s \,,\\
A_g\, x_{\gamma}+0.34 \,, \quad &  i=g \,,\end{array} \right. 
\label{eq:S_j}
\end{equation}
where $i$ is the parton flavor, $A_q=0.37$ and $A_g=0.19$ for a hard resolution scale of $p_T=5$ GeV.
Note that the model of Eq.\ (\ref{eq:S_j}) assumes no factorization breaking in the charm quark channel
since NLO QCD describes well diffractive photoproduction of open charm in $ep$ scattering (see above).

As one can see from Fig.\ \ref{fig:8} (top right), the resolved-only and interpolated schemes both lead indeed to a suppression at low $x_{\gamma}^{\rm obs.}$ that is twice as large as in the global suppression scheme. At intermediate values of $x_{\gamma}^{\rm obs.}$, the interpolated scheme is instead similar to global suppression, while for the pointlike region it is again similar to the resolved-only scheme. Since the distributions fall by two orders of magnitude from $x_{\gamma}^{\rm obs.}=0.85$ to 0.3, the differential cross section must be represented on a logarithmic scale and measurements at the EIC will require a high level of precision to distinguish between the different schemes. This should indeed be possible with the planned luminosities up two orders of magnitude larger than at HERA \cite{Skaritka:2018oxj}. The shape of the $\bar{p}_T$ distribution is also known to be sensitive to different schemes of factorization breaking 
\cite{Klasen:2008ah}, and this is also true for the global and resolved-only schemes at the EIC (Fig.\ \ref{fig:8}, top left). Interestingly, the interpolation scheme described above differs from the global scheme mostly in the larger normalization, which can be attributed to the fact that the cross section remains dominated by direct photons in the entire $\bar{p}_T$ range.
As expected, the distributions describing the momentum transfers from the proton to the pomeron (bottom left) and from the pomeron to the hard process (bottom right) have similar shapes for all three suppression schemes and differ again only in normalization.

\section{Diffraction on heavy nuclei}
\label{sec:4}


In the collider mode, it is rather straightforward to measure coherent diffraction on nuclei by selecting events with a large rapidity gap and requiring that no neutrons are produced in the zero-angle calorimeter (ZDC). Practically all events satisfying these requirements would correspond to coherent diffraction. However, measurements of the $t$-dependence would require the use of Roman pots at unrealistically small distances from the beam \cite{Frankfurt:2011cs}.

\subsection{Nuclear diffractive PDFs and nuclear shadowing}

Nuclear diffractive PDFs are defined similarly to those for nucleons as matrix elements of well-defined quark and gluon
operators between nuclear states with the condition that the final-state nucleus does not break, carries 
longitudinal momentum fraction $1-x_{\Pomeron}$, and that the four-momentum transfer squared is $t$.
%
%
%
%

As in the case of usual nuclear PDFs, nuclear diffractive PDFs are subject to nuclear modifications. In particular, at small
$x$ nuclear diffractive PDFs are expected to be suppressed compared to the coherent sum of free nucleon diffractive PDFs due to 
nuclear shadowing. In the model of leading twist nuclear shadowing \cite{Frankfurt:2011cs}, 
nuclear diffractive PDFs $f_{i/A}^{D}$ are obtained by summing a series corresponding to coherent diffractive scattering
on one, two, $\dots$, $A$ nucleons of the nuclear target, which gives in the small-$x_{\p}$ limit
\begin{equation}
f_{i/A}^{D}(z_{\p},Q^2,x_{\Pomeron})\approx 16 \pi B_{\rm diff} f_{i/p}^{D}(z_{\p},Q^2,x_{\Pomeron}) \int d^2 \vec{b} 
\left|\frac{1-e^{-\frac{A}{2}(1-i\eta)\sigma_{\rm soft}^i(x,Q^2)T_A(b)}}{(1-i\eta) \sigma_{\rm soft}^i(x,Q^2)}
\right|^2 \,.
\label{eq:nucl_diff}
\end{equation}
Here $B_{\rm diff} = 6$ GeV$^{-2}$ is the slope of the $t$-dependence of the $e p \to e^{\prime} Xp$ differential cross section
and $\eta=0.15$ is the ratio of the real to imaginary parts of the corresponding scattering amplitude;
$T_A(b)=\int dz \rho_A(b,z)$ is the optical nuclear density, where $\rho_A(b,z)$ is the nuclear density~\cite{DeJager:1987qc}
and $b$ is the transverse position (impact parameter) of the interacting nucleon in the nucleus;
$\sigma_{\rm soft}^i$ is an effective cross section controlling the strength of the interaction with target nucleons, which 
can be estimated using models of the hadronic structure of the virtual photon.
One can see from Eq.~(\ref{eq:nucl_diff}) that an account of nuclear shadowing leads in principle to an explicit violation of Regge factorization for nuclear diffractive PDFs.






A numerical analysis of Eq.~(\ref{eq:nucl_diff}) shows \cite{Frankfurt:2011cs} that the effect of nuclear shadowing 
in most of the kinematics only weakly
depends on flavor $i$, the momentum fractions $z_{\p}$ and $x_{\p}$, and the resolution scale $Q^2$.
Therefore, to a good approximation, one has the following relation
\begin{equation}
f_{i/A}^{D}(z_{\p},Q^2,x_{\Pomeron})\approx A R(x,A) f_{i/p}^{D}(z_{\p},Q^2,x_{\Pomeron}) \,,
\label{eq:nucl_diff_2}
\end{equation}
where $R(x,A) \approx 0.65$ is a weak function of $x$ and $A$ and is calculated using Eq.~(\ref{eq:nucl_diff}).

%
%


\subsection{NLO QCD predictions for the EIC}

Our predictions for the NLO QCD cross sections for coherent diffractive dijet photoproduction in $eA \to e^{\prime}+2\ {\rm jets}+ X+A$ 
scattering with different nuclear beams (U-238, Au-197, Cu-63, and C-12) at the EIC in our default set-up with $\sqrt{S}=92$ GeV are shown in Fig.\ \ref{fig:nucl}.
The cross sections are shown as functions of the jet average transverse momentum (top left), the jet rapidity difference (bottom left), the observed longitudinal
   momentum fractions of partons in the photon (top right) and pomeron (bottom right).
   As naturally follows from Eq.~(\ref{eq:nucl_diff_2}), the shapes of the nuclear cross sections repeat those for the 
   proton shown in Fig.\ \ref{fig:2}. 
   The free proton diffractive PDFs as parameterized in H1 2006 Fit B \cite{Aktas:2006hy}
   have been divided by a factor of 1.23 in order to take into account the fact that here
   we have no diffractive dissociation contributions, as the heavy nucleus is assumed to
   stay intact and no neutrons are assumed to be produced in the ZDC.

{\begin{figure}\centering
 \epsfig{file=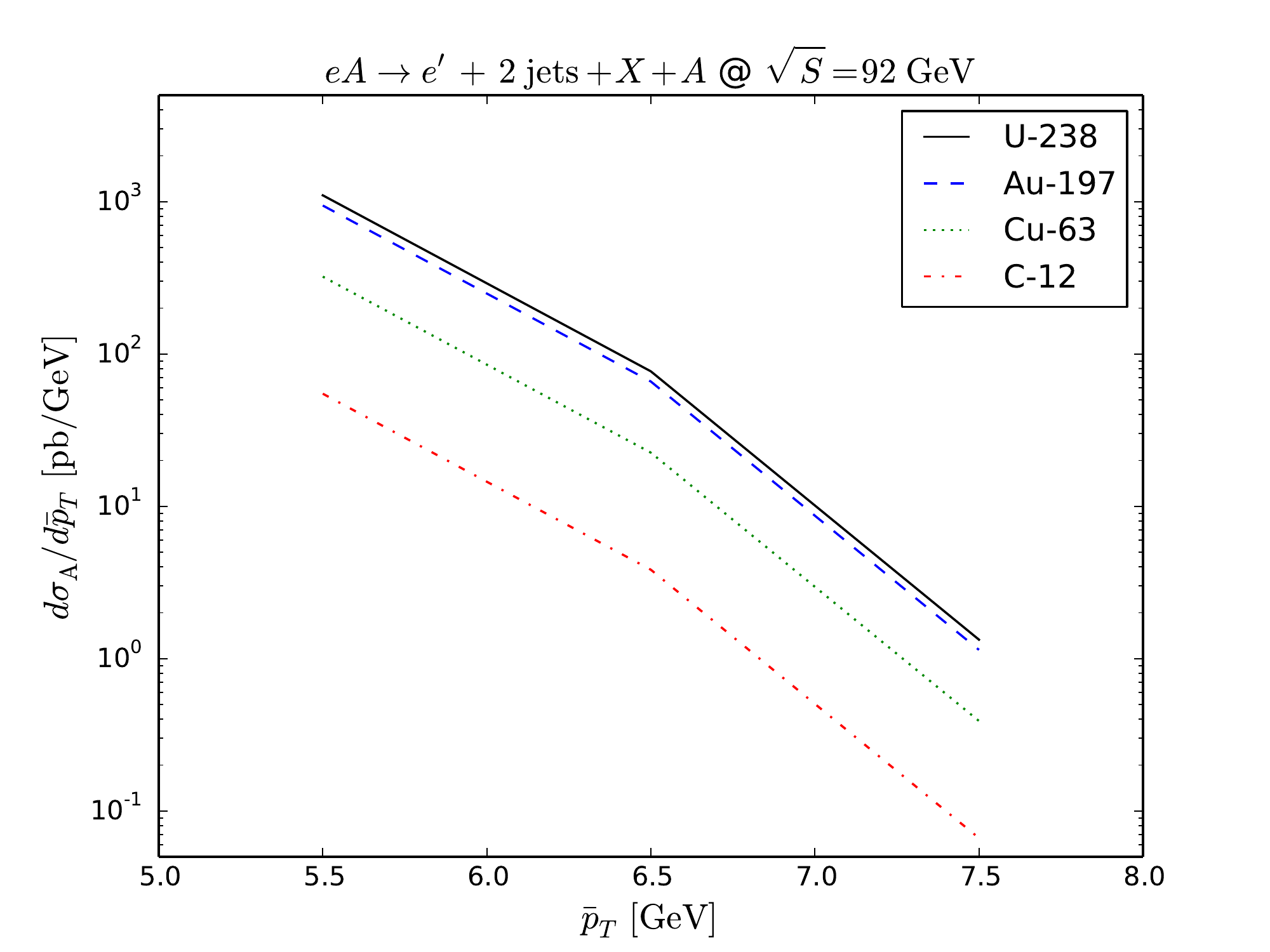,width=0.49\textwidth}
 \epsfig{file=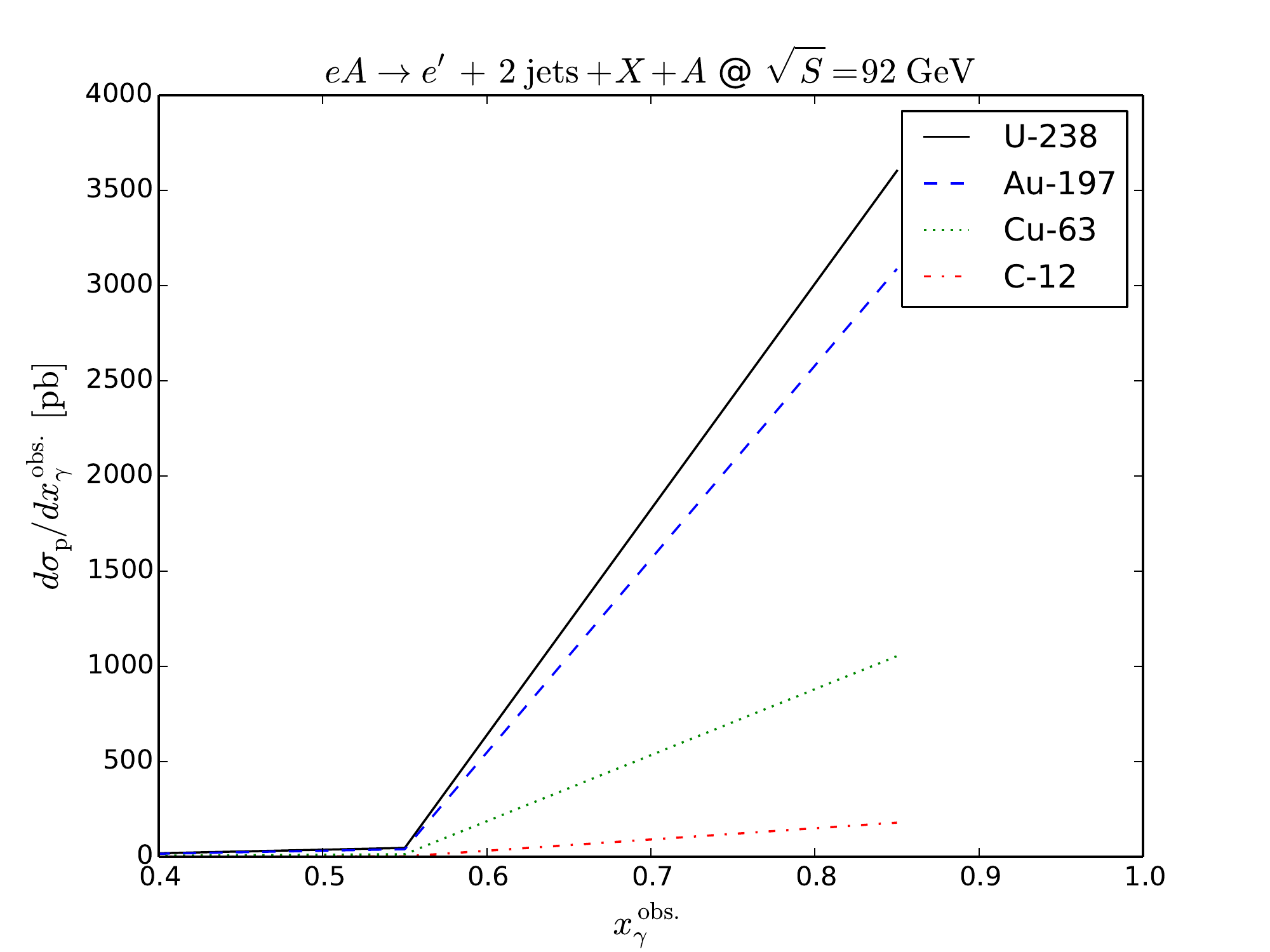,width=0.49\textwidth}
 \epsfig{file=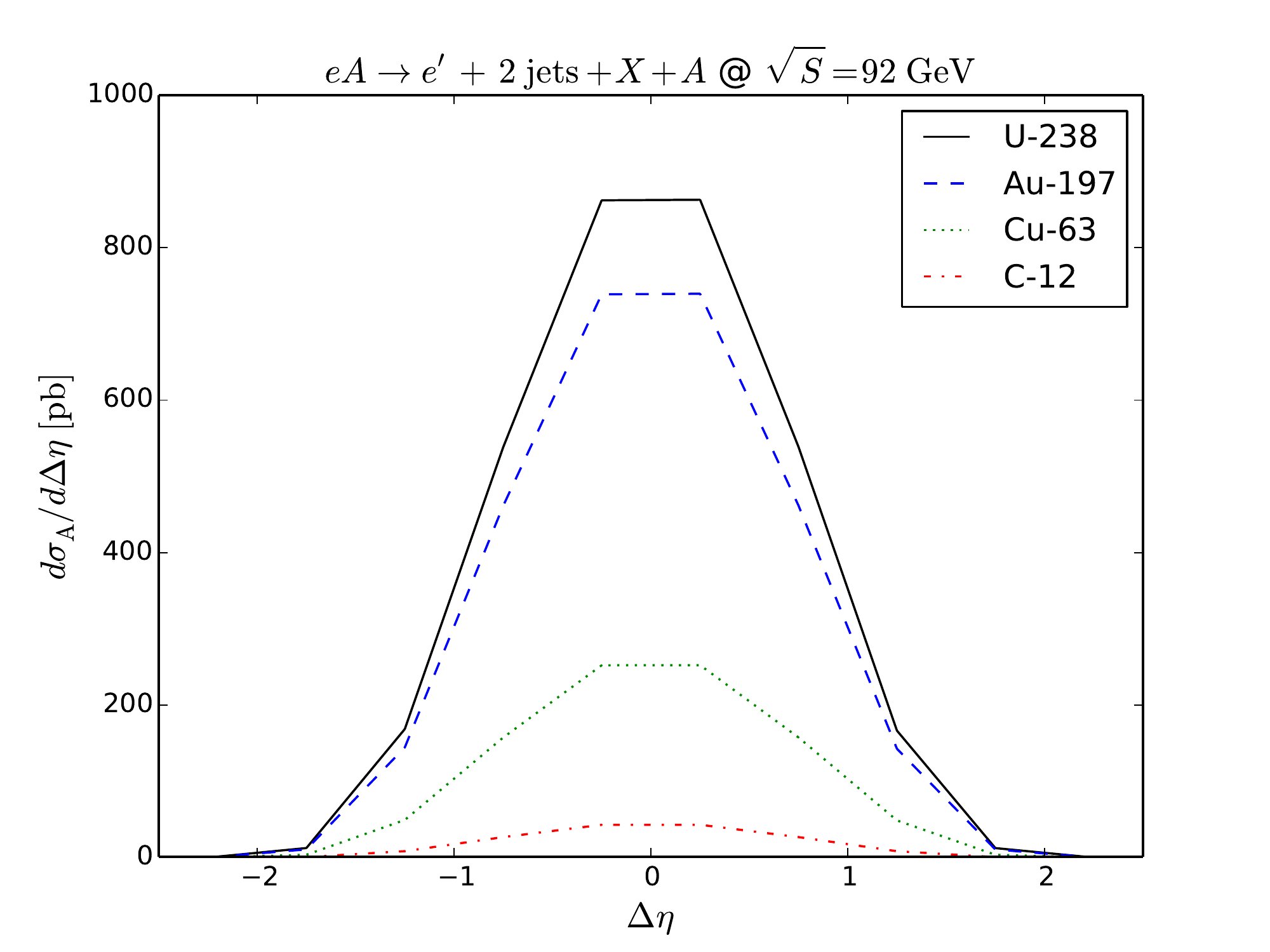,width=0.49\textwidth}
 \epsfig{file=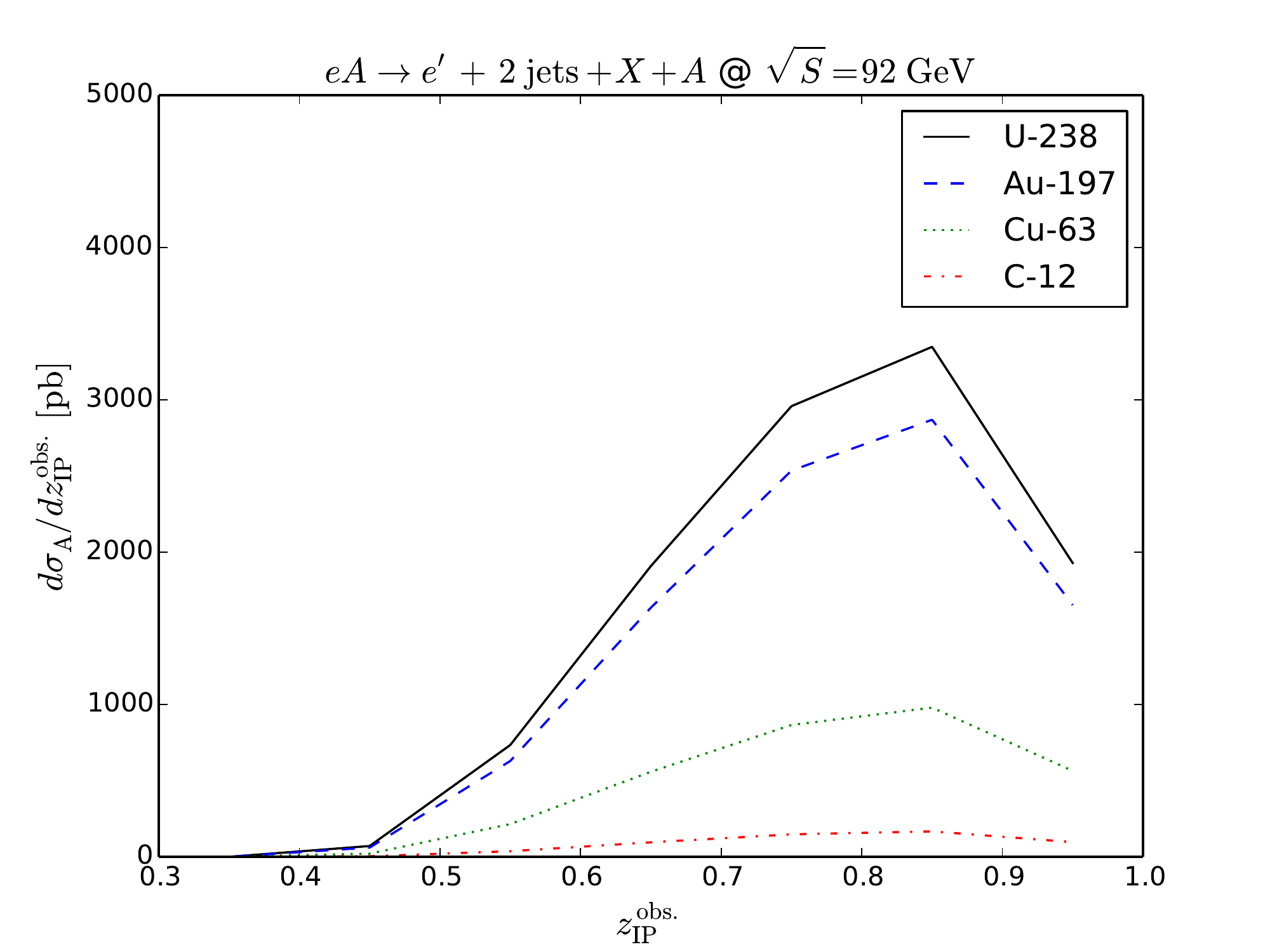,width=0.49\textwidth}
 \caption{NLO QCD cross sections for coherent  diffractive dijet photoproduction 
 $eA \to e^{\prime}+2\ {\rm jets}+ X+A$
 with various nuclear beams 
 and a center-of-mass energy per nucleon
 of $\sqrt{S}=92$ GeV at the EIC. The cross sections are shown as functions of 
 the jet average transverse momentum (top left) and rapidity difference (bottom left) 
 as well as the longitudinal
   momentum fractions in the photon (top right)
   and pomeron (bottom right).}
\label{fig:nucl}
\end{figure}}

%
%

\subsection{Factorization breaking}

As discussed in Sec.\ \ref{sunsec:fact_break}, collinear QCD factorization is violated in diffractive dijet photoproduction
due to soft inelastic interactions with the hadronic target. While the mechanism of this factorization breaking is not yet 
established, it is natural to expect that the effect will be more pronounced for nuclear targets since
 the gap survival probability is significantly smaller for nuclei than for the proton. 
For instance, using the commonly used two-state eikonal model \cite{Khoze:2000wk,Kaidalov:2003xf}, one estimates\  \cite{Guzey:2016tek} that the suppression factor in the relevant energy range
 is $S^2=0.4$ for the proton and $S^2=0.04$ for heavy nuclei.
 
 Figure \ref{fig:nucl_fact} shows our predictions for the NLO QCD cross section for coherent 
 diffractive dijet photoproduction with Au-197 beams at the EIC. For heavy nuclei, we 
 unfortunately cannot enhance the resolved-photon contribution to increase the differences between
 competing factorization breaking schemes by assuming a higher beam energy of 275 GeV as for
 protons, but we remain limited to 100 GeV per nucleon. We can, however, perform our study for
 the larger range in $x_{\p} <0.10$ instead of 0.03, which, as we have seen in Fig.\
 \ref{fig:6}, also increases the accessible range in $x_\gamma$. In Fig.\ \ref{fig:nucl_fact},
 the solid lines correspond to the case where we apply a global suppression factor of $S^2=0.5$ to our calculated cross sections
 as in the proton case, while the dotted lines correspond to the factorization breaking scenario, where only the resolved-photon
 contribution is suppressed, now by a factor of $S^2=0.04$ (see the discussion above).
 A comparison of these two schemes shows that due to the dominance of the direct photon contribution in most of the 
 EIC kinematics, they lead to similar shapes of the kinematic distributions that differ only in normalization, with the important exception of the $x_\gamma$-distribution (top right), which
below values of 0.5 is negligible for resolved-only suppression and has potentially measurable support only in the global suppression scheme.
 
{\begin{figure}\centering
 \epsfig{file=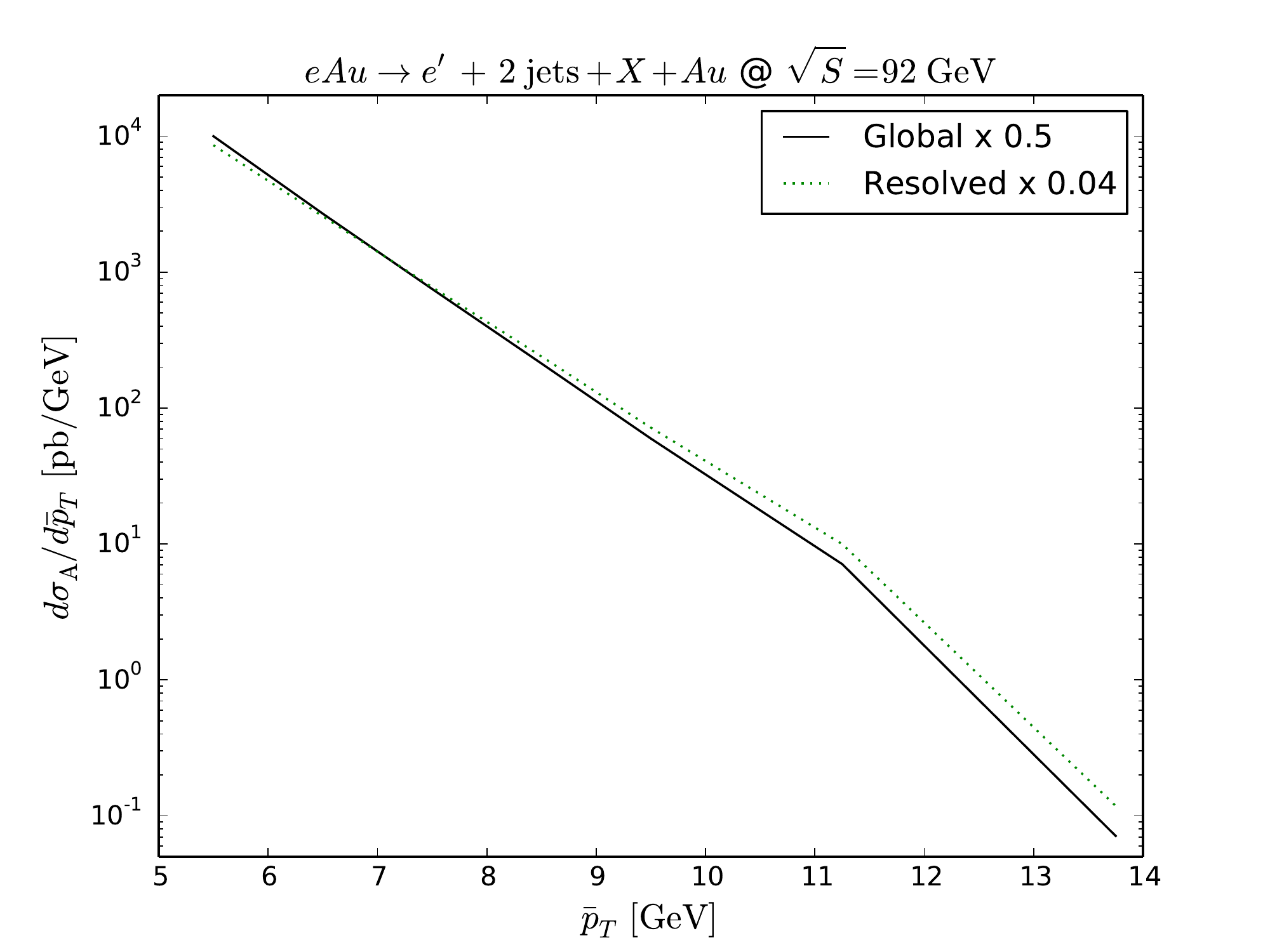,width=0.49\textwidth}
 \epsfig{file=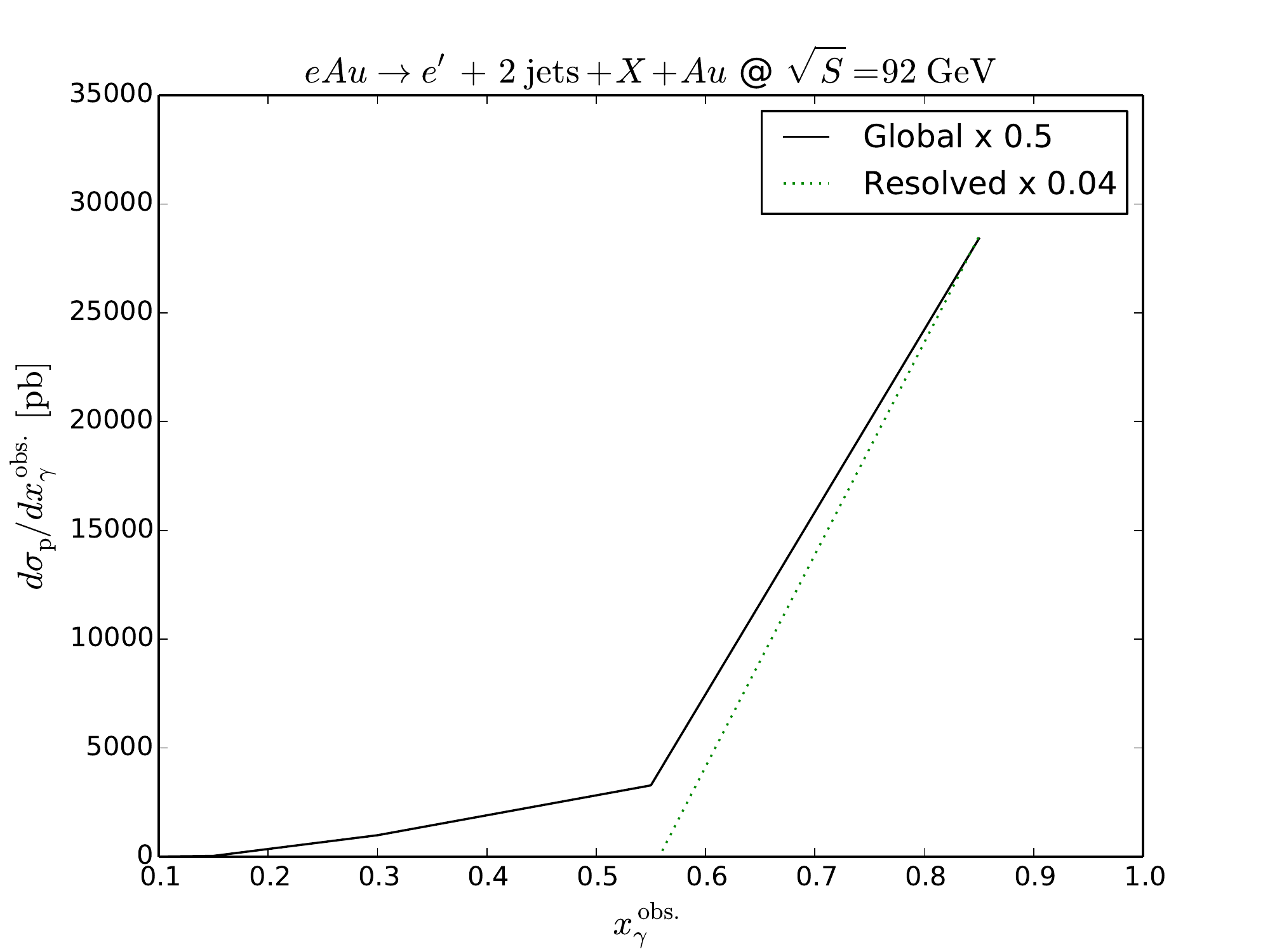,width=0.49\textwidth}
 \epsfig{file=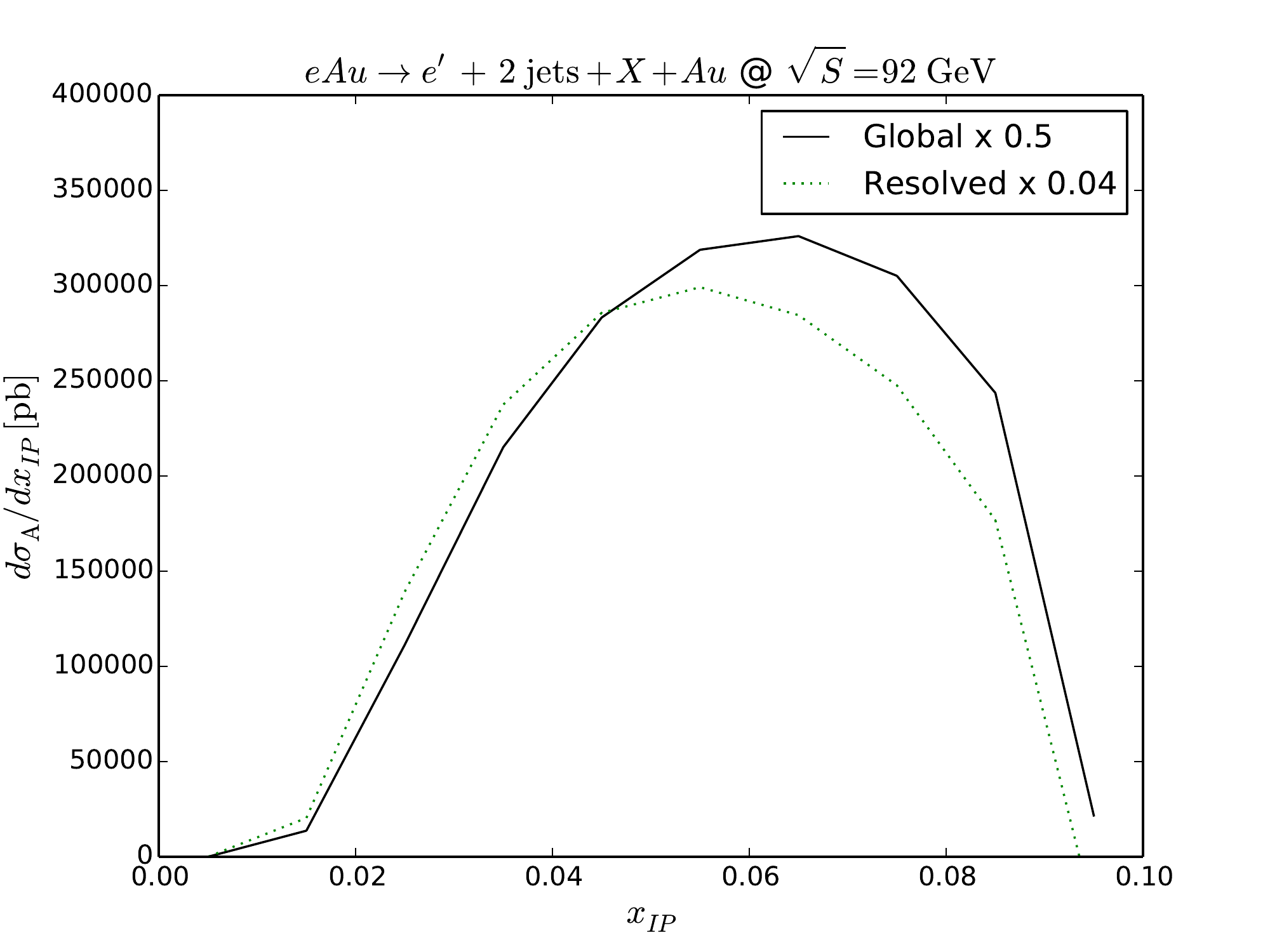,width=0.49\textwidth}
 \epsfig{file=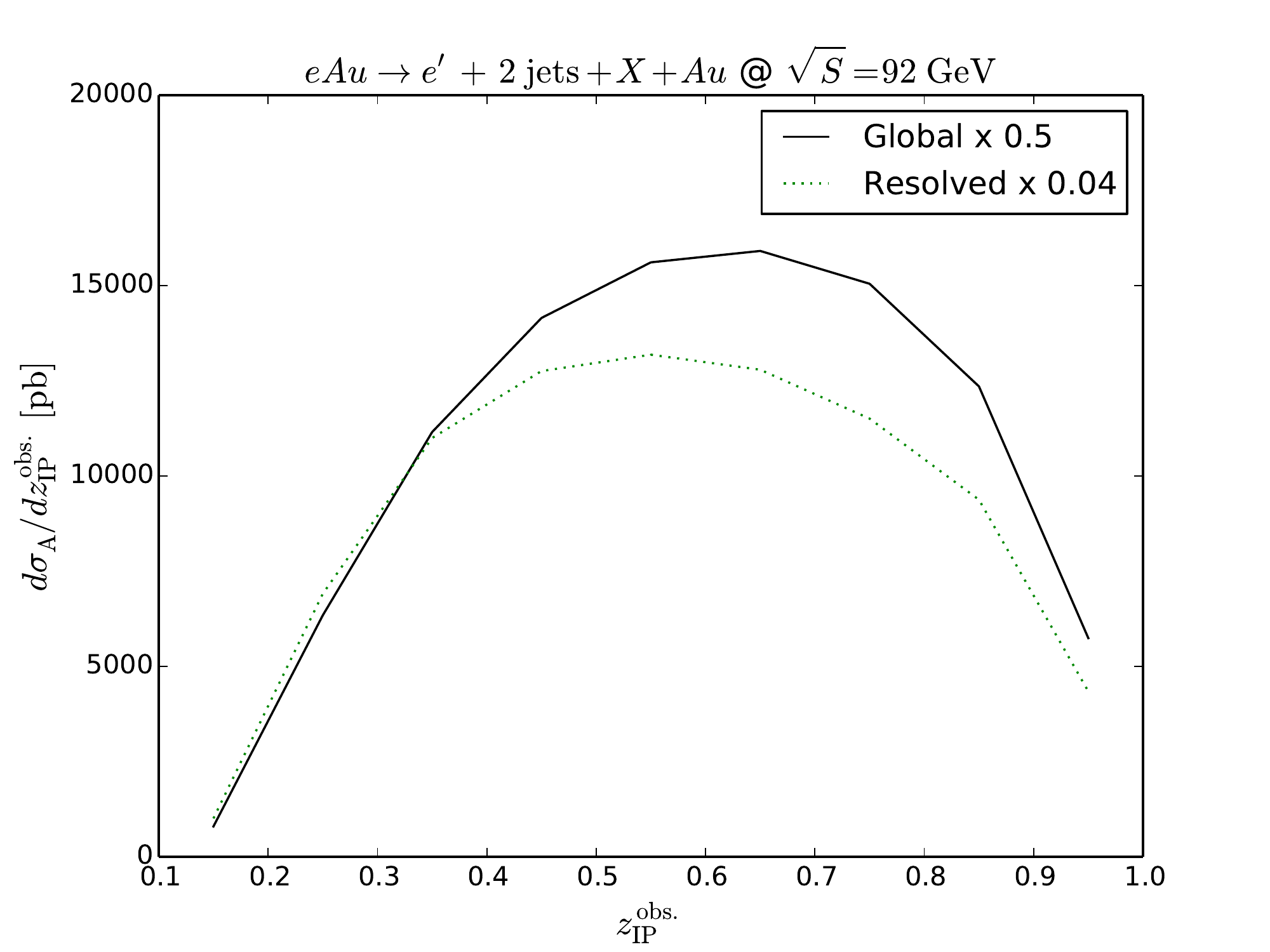,width=0.49\textwidth}
 \caption{NLO QCD cross sections for coherent diffractive dijet photoproduction 
   $eAu \to e^{\prime}+2\ {\rm jets}+ X+Au$ at $\sqrt{S}=92$ GeV at the EIC
   with an extended range in $x_{\p}  <0.1$. Shown are distributions in the jet average
   transverse momentum (top left) as well as the (observed) longitudinal
   momentum fractions of the photon (top right), the pomeron (bottom left)
   and the partons in the pomeron (bottom right). We compare two different schemes of
   factorization breaking, i.e.\ global suppression by a factor of 0.5 (full black)
   with only resolved-photon suppression by a factor of 0.04 (dotted green curves).}
\label{fig:nucl_fact}
\end{figure}}

\section{Conclusion}
\label{sec:5}

To summarize, we have in this paper presented a first and extensive study of
diffractive dijet photoproduction at the recently approved EIC. Using our established
formalism of NLO QCD calculations, we have illuminated various aspects of this
interesting scattering process. We started by determining the cross sections
to be expected in the most important differential distributions as well as the size
of the NLO corrections. We then discussed the sensitivity to pomeron PDFs as
a function of momentum fraction and resolution scale as well as the contribution
from the higher reggeon tractectory. One of our two main results is that the EIC
has the potential
to address conclusively the mechanism of factorization breaking, but that this
will require a high proton beam energy and/or a large longitudinal momentum transfer
from the proton/nucleus to
the pomeron. Then the question will hopefully be answered whether and to what extent
factorization breaking occurs globally in photoproduction or whether only the resolved
photon contribution or some of its components (light/heavy quark-antiquark pairs,
VMD contributions) are suppressed. Our second main result comprises predictions
for diffractive dijet photoproduction on nuclei, which might -- perhaps for the first
time -- give access to nuclear diffractive PDFs. Here, we made numerical predictions
for four different nuclei, ranging from carbon to uranium, as well as again for
different factorization breaking schemes.

As an outlook, let us point out that at HERA also dijet production with leading neutrons
has been studied \cite{Aktas:2004gi,Breitweg:2000nk}. These processes have been interpreted
in terms of virtual charged-pion exchanges and gave first information on the structure of
(virtual) pions at previously unaccessible values of $x$ \cite{Klasen:2001sg}. It would be
very interesting to continue these studies at the EIC, also in view of possible 
factorization breaking in these processes \cite{Klasen:2006hs}, and perhaps even extend them
to dissociation processes in collisions with heavy nuclei.



\section*{Acknowledgements}
\noindent

The authors thank Wim Cosyn and the other organizers of the 1st EIC Yellow
Report workshop at Temple University for stimulating this study and Paul Newman
for his interesting questions, in particular on the impact of a larger range
in $x_{\p}$. Financial support by the DFG through the grant KL 1266/9-1 within
the framework of the joint German-Russian project ``New constraints on nuclear
parton distribution functions at small $x$ from dijet production in $\gamma A$
collisions at the LHC" is gratefully acknowledged. VG's research is also
supported in part by RFBR, research project 17-52-12070.

\end{document}